\documentclass[%
preprint,
superscriptaddress,
amsmath,amssymb,
aps,longbibliography,
]{revtex4-1}

\usepackage{graphicx}
\usepackage{verbatim}
\usepackage[version=4]{mhchem}
\usepackage{xcolor}
\usepackage{physics}
\usepackage[normalem]{ulem}
\usepackage{pdfpages}
\usepackage{pgffor}
\usepackage{outlines}
\usepackage{booktabs}
\usepackage{multirow}
\usepackage{comment}

\makeatletter
\AtBeginDocument{\let\LS@rot\@undefined}
\makeatother

\begin{document}

\title{Comparing the latent features of universal \\ machine-learning interatomic potentials}


\author{Sofiia Chorna}
\affiliation{Laboratory of Computational Science and Modeling, Institut des Mat\'eriaux, \'Ecole Polytechnique F\'ed\'erale de Lausanne, 1015 Lausanne, Switzerland}

\author{Davide Tisi}
\affiliation{Laboratory of Computational Science and Modeling, Institut des Mat\'eriaux, \'Ecole Polytechnique F\'ed\'erale de Lausanne, 1015 Lausanne, Switzerland}
\affiliation{Department of Energy Conversion and Storage, Technical University of Denmark, 2800 Kongens Lyngby, Denmark}

\author{Cesare Malosso}
\affiliation{Laboratory of Computational Science and Modeling, Institut des Mat\'eriaux, \'Ecole Polytechnique F\'ed\'erale de Lausanne, 1015 Lausanne, Switzerland}

\author{Wei Bin How}
\affiliation{Laboratory of Computational Science and Modeling, Institut des Mat\'eriaux, \'Ecole Polytechnique F\'ed\'erale de Lausanne, 1015 Lausanne, Switzerland}

\author{Michele Ceriotti}
\email{michele.ceriotti@epfl.ch}
\affiliation{Laboratory of Computational Science and Modeling, Institut des Mat\'eriaux, \'Ecole Polytechnique F\'ed\'erale de Lausanne, 1015 Lausanne, Switzerland}

\author{Sanggyu Chong}
\email{sanggyu.chong@epfl.ch}
\affiliation{Laboratory of Computational Science and Modeling, Institut des Mat\'eriaux, \'Ecole Polytechnique F\'ed\'erale de Lausanne, 1015 Lausanne, Switzerland}

\date{\today}

\begin{abstract}
The past few years have seen the development of ``universal'' machine-learning interatomic potentials (uMLIPs) capable of approximating the ground-state potential energy surface across a wide range of chemical structures and compositions with reasonable accuracy. While these models differ in the architecture and the dataset used, they share the ability to compress a staggering amount of chemical information into descriptive latent features. Herein, we systematically analyze what the different uMLIPs have learned by quantitatively assessing the relative information content of their latent features with feature reconstruction errors, and observing how the trends are affected by the choice of training set and training protocol. We find that uMLIPs encode the chemical space in significantly distinct ways, with substantial cross-model feature reconstruction errors. When variants of the same model architecture are considered, trends become dependent on the dataset, target, and training protocol of choice. We also observe that fine-tuning of a uMLIP retains a strong pre-training bias in the latent features. Finally, we discuss how atom-level features, which are directly output by MLIPs, can be compressed into global structure-level features via concatenation of progressive cumulants, each adding significantly new information about the variability across the atomic environments within a given system.
\end{abstract}
\maketitle

\section{Introduction}

Machine learning (ML) has brought transformative changes to the field of computational materials science. Data-driven algorithms enable simulations and electronic structure calculations with near \textit{ab initio} accuracy to reach unprecedented length and time scales \cite{Jacobs2025, Schmidt2019MLinMaterials, Qu2024, Mahmoud2020, How2025, shnorb, Suman2025}.
A particularly active frontier in this field is the development of ``universal'' machine-learned interatomic potentials (uMLIPs) that aim for applicability across the entire chemical composition space~\cite{DPA3, UMA,  MACE, mace_multihead, gnome, PET-MAD-2025, MatterSim, OpenLAMv1}. Such models alleviate the need for laborious dataset construction and model training from scratch, and instead allow zero-shot deployment or fine-tuning of existing uMLIPs, further accelerating applications across diverse areas of chemistry and materials science.

To date, several dozens of uMLIPs have been developed, each employing distinct atomic descriptors, neural network (NN) architectures, and training datasets of varying scope and composition~\cite{MPTRAJ, Alexandria, MAD, MATPES, SPICE, OC20, ODAC, omat24, OMC25, OMOL}. Despite their rapidly growing diversity, most uMLIPs have been evaluated primarily on curated benchmarking suites that assess their accuracy across standard material modeling tasks \cite{matbench, LAMBench, chiang2025mliparena}. This limited perspective motivates a deeper examination of the similarities and differences among uMLIPs, particularly how they represent and interpret the chemical space \cite{muller2025peeringinside}.

Regardless of their architectural details, all uMLIPs share a key characteristic: they encode the multidimensional description of atomic systems, comprised of types and positions of constituent atoms, into a compact latent representation containing only a few hundred features. 
These latent representations embody the distinct ways in which different uMLIPs perceive and organize the chemical space, and the utility of these latent features has been demonstrated in various contexts: uncertainty quantification \cite{llpr, fast_uncertainty, uncertainty_nn, kulichenko2024uncertainty, Janet2019ChemSci, Feldmann2025}, outlier detection \cite{Vazquez-Salazar2025, Schwalbe_Koda_2025}, active learning \cite{Xie2021BayesianFF, Vandermause2020, kulichenko2024uncertainty}, dataset curation \cite{kulichenko2024data, zou2025data, Glielmo2022}, and system classification \cite{de+16jci, stability_classification}.

In this work, we leverage the latent features of uMLIPs to systematically and quantitatively analyze the similarities and differences in how these models perceive and structure the chemical space. While recent complementary works have focused on discerning whether there exists a converged or aligned representation of matter across the uMLIPs \cite{bombarelli2025, li2025platonicrepresentation}, here we adopt a purely statistical approach, using the global and local feature reconstruction errors of Goscinski \textit{et al.}~\cite{Goscinski_2021} to assess the information content across the latent features of four uMLIPs: MACE-MP-0b3 \cite{MACE}, PET-MAD \cite{PET-MAD-2025}, DPA-3.1 \cite{DPA3}, and UMA-S-1P1 \cite{UMA}. We investigate the mutual reconstructability of the uMLIP latent features, as well as those of model variants trained with different datasets and training strategies. We also extend the evaluation to the latent features of PET-MAD and those of PET-MAD-DOS \cite{PET-MAD-DOS-2025}, a model with nearly identical architecture and dataset but trained for a different target property. We then study the evolution of the latent space during fine-tuning, and explore how the atomic latent features can be efficiently aggregated into global features from the information content standpoint.

We find that all uMLIPs uniquely encode chemical space with significant cross-model reconstruction errors, where some models are more accurately linearly reconstructed by others. When considering the variants of model architectures targeting different datasets, we observe that single-task models (MACE-MP-0b3) and multi-task models (MACE-mh-1 \cite{mace_multihead}, DPA-3.1) share relatively consistent representations across datasets, while the mixture of linear experts (MoLE) model (UMA-S-1P1) specializes the features far more distinctly. We observe that the information content of ``last-layer'' and ``backbone'' features in a model architecture is similar, yet the backbone features encode more information that last-layer features cannot reconstruct. We also demonstrate that fine-tuning of the uMLIP exhibits a strong pre-training bias in the latent feature space. Finally, we demonstrate the compression of local, atom-level features into global, structure-level features by concatenating their progressive cumulants, resulting in an information-rich structural descriptor that obeys desirable scaling laws and preserves the knowledge of configurational inhomogeneity. 

\section{Methods}

\begin{table*}[tbh]
    \centering
    \caption{Key characteristics of the uMLIPs considered in this work. Dataset sizes are reported by the number of structures. For UMA-S-1P1, 6M denotes the active parameters used per structure during inference, while 150M is the total across all the experts.}
    \begin{tabular}{|p{3.4cm}|p{3.1cm}|p{3.1cm}|p{3.1cm}|p{3.1cm}|} 
        \hline
          & {\centering MACE-MP-0b3\par} 
          & {\centering PET-MAD\par}
          & {\centering DPA-3.1\par}
          & {\centering UMA-S-1P1\par} \\
        \hline

        \textbf{Training strategy} &
        {\centering Single model\par} &
        {\centering Single model\par} &
        {\centering Multi-task model\par} &
        {\centering Mixture of linear experts model\par} \\

        \textbf{Dataset(s)} &
        {\centering MPtrj (1.58M) \cite{MPTRAJ}\par} &
        {\centering MAD (0.096M) \cite{MAD}\par} &
        {\centering OpenLAM-v1 (163M) \cite{OpenLAMv1}\par} &
        {\centering OMat24, OC20, ODAC25, OMC25, OMol25 (total 569M) \cite{omat24, OC20, ODAC, OMC25, OMOL}\par} \\

        \textbf{Equivariance} &
        {\centering E(3)-equivariant\par} &
        {\centering Rotationally unconstrained; equivariance learned by data augmentation\par} &
        {\centering SE(3)-invariant\par} &
        {\centering E(3)-equivariant\par} \\

        \textbf{\# of parameters} &
        {\centering 4.69M\par} &
        {\centering 3.3M\par} &
        {\centering 3.26M\par} &
        {\centering 6M (active), 150M (total)\par} \\

        \textbf{Feature dim.} &
        {\centering 144\par} &
        {\centering 512\par} &
        {\centering 240\par} &
        {\centering 128\par} \\

        \textbf{Cutoff (\AA)} &
        {\centering 6.0\par} &
        {\centering 4.5\par} &
        {\centering 6.0\par} &
        {\centering 6.0\par} \\
        \hline
    \end{tabular}
    \label{tab:umlips}
\end{table*}

For our analysis, we mainly consider MACE-MP-0b3, PET-MAD, DPA-3.1, and UMA-S-1P1, which constitute a non-exhaustive yet diverse set of uMLIPs that differ vastly in their message-passing NN architectures, learning strategies, and training datasets (see Table \ref{tab:umlips}). MACE-MP-0b3 \cite{MACE} is built on an E(3)-equivariant message-passing NN that constructs higher body-order correlations using equivariant tensor products and body-ordered polynomials. PET-MAD \cite{PET-MAD-2025} is based on a rotationally unconstrained transformer-based graph neural network (GNN), where strict geometrical equivariance is not enforced by architecture but is effectively learned through rotational data augmentation during training \cite{PET}. DPA-3.1 \cite{DPA3} is an attention-based model of the Deep Potential (DP) family \cite{DeepMD-kit-v3} that produces fully rotation- and translation-invariant atomic descriptors via a gated attention mechanism. UMA-S-1P1 \cite{UMA} is implemented upon a large-scale E(3)-equivariant mixture-of-experts architecture built on the eSEN \cite{eSEN} backbone that uses a MoLE design. For UMA-S-1P1 and DPA-3.1, which are a MoLE model and a multi-head model, respectively, we use the expert or the head corresponding to the OMat24 dataset \cite{omat24}, given that MACE-MP-0b3 and PET-MAD have been trained on similar materials-oriented datasets.

We define the latent features of each uMLIP as the tokens extracted from a given layer of the model architecture. These features have been processed through multiple layers of message-passing to encode the information aggregated from neighboring atoms within the defined cutoff radius. We primarily focus on the last-layer features, corresponding to the representation most refined for the regression task obtained after all message-passing iterations and subsequent multilayer perceptron (MLP) transformations. Since all considered uMLIPs adopt a locality ansatz, we extract atomic features $\boldsymbol{\xi}(A_i)^{F}$ for each atom $i$ in structure $A$ via a forward-pass through the model $F$.

To enable a consistent, scale-invariant comparison across the models, the features are centered to zero mean and globally scaled so that their total variance across all dimensions equals one on the analysis dataset for each uMLIP, thereby preserving the relative scales between feature dimensions.

\subsection{Feature space comparison metrics}
\label{sec:reconstruction-errors}

In our analysis, we primarily take a \textit{quantitative} approach in studying the similarities and differences between the latent space of uMLIPs. To this end, we compute the global feature reconstruction error (GFRE) and the local feature reconstruction error (LFRE) proposed by Goscinski \textit{et al.}~\cite{Goscinski_2021}. These metrics assess the relative expressive power of feature spaces by determining how well one set of features can reconstruct another. 

Let $\mathcal{D}$ be a set of $n$ atomic environments. For two models $F$ and $F'$, their feature matrices are first computed:

\begin{equation}
X^F_{\mathcal{D}} = 
\begin{bmatrix}
\boldsymbol{\xi}_1^{F} \\
\boldsymbol{\xi}_2^{F} \\
\vdots \\
\boldsymbol{\xi}_n^{F}
\end{bmatrix}
\in \mathbb{R}^{n \times d_F}, 
\qquad
X^{F'}_{\mathcal{D}} =
\begin{bmatrix}
\boldsymbol{\xi}_1^{F'} \\
\boldsymbol{\xi}_2^{F'} \\
\vdots \\
\boldsymbol{\xi}_n^{F'}
\end{bmatrix}
\in \mathbb{R}^{n \times d_{F'}},
\end{equation}
where \(d_F\) and \(d_{F'}\) denote the dimensionalities of the feature vectors for models \(F\) and \(F'\), respectively, which do not need to be equal.

To compute the GFRE, we determine the optimal linear mapping $\widehat{P}_{FF'} \in \mathbb{R}^{d_F \times d_{F'}}$ that minimizes the reconstruction loss over a training set $\mathcal{D}_{\mathrm{train}}$:
\begin{equation}
\widehat{P}_{FF'} = \arg\min_{\widehat{P}} \left\| X_{\mathcal{D}_{\mathrm{train}}}^{F'} - \widehat{P} X_{\mathcal{D}_{\mathrm{train}}}^F \right\|^2_F, 
\end{equation}
where $\|\cdot\|_F$ denotes the Frobenius norm that measures the overall square differences between two matrices, i.e., total Euclidean distances between all pairs of atomic feature vectors.

After obtaining the optimized $\widehat{P}_{FF'}$, the GFRE is computed for $\mathcal{D}_{\mathrm{test}}$ as the root mean square error (RMSE) between $X_{\mathcal{D}_{\mathrm{test}}}^{F'}$ and $\widehat{P}_{FF'} X_{\mathcal{D}_{\mathrm{test}}}^F$:
\begin{equation}
\text{GFRE}_{\mathcal{D}_{\mathrm{test}}}(F, F') = \sqrt{ \frac{ \left\| X_{\mathcal{D}_\text{test}}^{F'} - \widehat{P}_{FF'}  X_{\mathcal{D}_\text{test}}^F \right\|^2_F }{ |\mathcal{D}_\text{test}| } },
\end{equation}
where $ |\mathcal{D}_\text{test}|$ is the total number of atomic environments in the test set $\mathcal{D}_\text{test}$.
A low GFRE indicates that the latent features of $F$ contain sufficient information to linearly reconstruct the latent features of $F'$, meaning that both models encode similar information about the atomic environments. In contrast, a large value signifies that features of $F'$ carry supplementary information that cannot be recovered as a linear transformation of the features of $F$. The GFRE mapping is directional, i.e., $\text{GFRE}(F, F') \neq \text{GFRE}(F', F)$.

As the GFRE exclusively targets linear transformations between the features, it overlooks potential nonlinear mapping between the feature spaces.
The LFRE is an alternative metric that indirectly addresses this limitation: it quantifies how well a \emph{local} neighborhood in one feature space can be reconstructed from its corresponding neighborhood in another.

For each test environment $i \in \mathcal{D}_\text{test}$, we identify its k nearest neighbors in the training set $\mathcal{D}_\text{train}$ according to the Euclidean distance in feature space F. Collectively, these neighbor indices are $\mathcal{N}_i^F = \{j_1, \dots, j_k\} \subset \mathcal{D}_\text{train}$. For these neighbors, we denote source-space and target-space features as:
\begin{equation}
X^F_{\mathcal{N}_i^F} = 
\begin{bmatrix}
\boldsymbol{\xi}_{j_1}^{F} \\
\boldsymbol{\xi}_{j_2}^{F} \\
\vdots \\
\boldsymbol{\xi}_{j_k}^{F}
\end{bmatrix}
\in \mathbb{R}^{k \times d_F}, 
\qquad
X^{F'}_{\mathcal{N}_i^{F}} =
\begin{bmatrix}
\boldsymbol{\xi}_{j_1}^{F'} \\
\boldsymbol{\xi}_{j_2}^{F'} \\
\vdots \\
\boldsymbol{\xi}_{j_k}^{F'}
\end{bmatrix}
\in \mathbb{R}^{k \times d_{F'}}.
\end{equation}

\noindent To restrict the learning problem to modeling only relative displacements on the local manifold patch, each neighborhood is centered by subtracting its mean:
\begin{equation}
\widetilde{X}_{\mathcal{N}_i^F}^{F}
= X_{\mathcal{N}_i^F}^{F} - \bar{\boldsymbol{\xi}}_{\mathcal{N}_i^F}^{F},\qquad
\widetilde{X}_{\mathcal{N}_i^F}^{F'}
= X_{\mathcal{N}_i^F}^{F'} -
\bar{\boldsymbol{\xi}}_{\mathcal{N}_i^F}^{F'}.
\end{equation}

\noindent A local linear map is then fitted to align the centered neighborhood of these $k$ training points from $F$ to $F'$:
\begin{equation}
\widehat{P}^{FF'}_{i} = \arg\min_{\widehat{P}} \left\| \widetilde{X}_{\mathcal{N}_i^F}^{F'} - \widehat{P} \widetilde{X}_{\mathcal{N}_i^F}^{F} \right\|_F^2.
\end{equation}

\noindent This local neighborhood mapping is then applied to reconstruct the centered feature of the test point $i$ itself:
\begin{equation}
\widehat{\boldsymbol{\xi}}_i^{F'} =
\bar{\boldsymbol{\xi}}_{\mathcal{N}_i^F}^{F'}
+ \widehat{P}^{FF'}_{i}
(\boldsymbol{\xi}_i^{F} -
\bar{\boldsymbol{\xi}}_{\mathcal{N}_i^F}^{F}
).
\end{equation}

\noindent Finally, the LFRE is the RMSE over all test environments:
\begin{equation}
\text{LFRE}_{\mathcal{D}_{\mathrm{test}}} (F, F') = \sqrt{ \frac{ \sum_{i \in \mathcal{D}_{\mathrm{test}}} 
\left\|\boldsymbol{\xi}_i^{F'} - \widehat{\boldsymbol{\xi}}_i^{F'}
\right\|_2^2
}{|\mathcal{D}_{\mathrm{test}}|}}.
\end{equation}

\noindent The neighborhood size $k$ can be optimized, but here we set it to the smaller feature dimension between the two models for simplicity. Low LFRE values show that the local structure of $F'$ can be reconstructed by $F$, hence the two feature spaces contain similar information, even if they differ globally. 

We clarify that the definitions of ``global'' and ``local'' in the context of feature reconstruction errors do not correspond to the usual atomic versus system-wide considerations in atomistic ML (see SI \ref{sec:si-errors} for a schematic overview). Rather, ``global'' denotes a single linear transformation applied across the entire dataset of atomic features, whereas ``local'' involves fitting separate linear mappings for localized neighborhoods within the high-dimensional feature space. Both metrics can operate on both atomic-level and structure-level features, but differ in how they assess reconstructability across the feature manifold.

Owing to the feature normalization, both the GFRE and LFRE are usually bounded between $0$ and $1$, with $0$ signifying perfect linear reconstruction, thus comparable information between the feature spaces. However, the upper bounds can be exceeded in case of reconstruction failure due to ill-conditioned linear mappings, which indicates large dissimilarity or incompatibility between the feature spaces. For our analysis, we use the GFRE and LFRE implementations in \textit{scikit-matter} \cite{scikit-matter} with ridge regression and default 2-fold cross-validation to determine the ridge regularization, with regularization $\alpha$ values scanned over a logarithmic range from $10^{-9}$ to 1, and the final value was selected as the smallest $\alpha$ that provided numerical stability.

\subsection{Feature space projection}

In a few examples, we complement this quantitative analysis by a visual mapping of the uMLIP features to a lower-dimensional space using Principal Covariates Regression (PCovR) \cite{pcovr, cersonsky_pcovr}.
This dimensionality reduction technique interpolates between two objective functions: maximizing variance retention in the original features via principal component analysis (PCA), and optimizing linear regression performance on a target property.
In our context, by mixing in the regression term in addition to the PCA term, we gently enforce a degree of consistency across the maps produced for different models to aid interpretation.

Given a set of high-dimensional feature matrix \(X \in \mathbb{R}^{n \times d}\) and target property matrix \(Y \in \mathbb{R}^{n \times p}\), PCovR determines a latent space projection $\mathbf{T} \in \mathbb{R}^{n \times k}$ that minimizes the loss:
\begin{equation}
\mathcal{L} = \alpha\| X - \widehat{X} \|^2_F +  (1 - \alpha) \| Y - \widehat{Y} \|^2_F,
\end{equation}

\noindent where $\widehat{X} = \mathbf{T} P_{X}^{\top}$ is the approximation of the original feature space $X$ (PCA term) and $\widehat{Y} = \mathbf{T} P_{Y}^{\top}$ is the prediction of targets (linear regression term). The matrices $P_{X}$ and $P_{Y}$ represent projection weights that map latent coordinates back to the original feature and target dimensions. The interpolation between the two terms is controlled by a tunable parameter $\alpha \in \left[0, 1\right]$. In our analysis, we use \textit{scikit-matter} implementation with $\alpha=0.5$ and $k = 2$ principal components.

To prepare the inputs, atomic features are aggregated to the structure-level by computing the cumulants introduced in Sec.~\ref{sec:cumulants} across atoms within each structure.
The target property $Y$ is defined as the cohesive energies per atom, calculated by subtracting an atomic baseline energy from the reference total potential energy of each structure and normalizing by the number of atoms.

\section{Results}

\subsection{Latent feature comparison between uMLIPs}
\label{subsec:errors-umlip}

\begin{figure}[htbp]
    \centering
    \includegraphics[width=\linewidth]{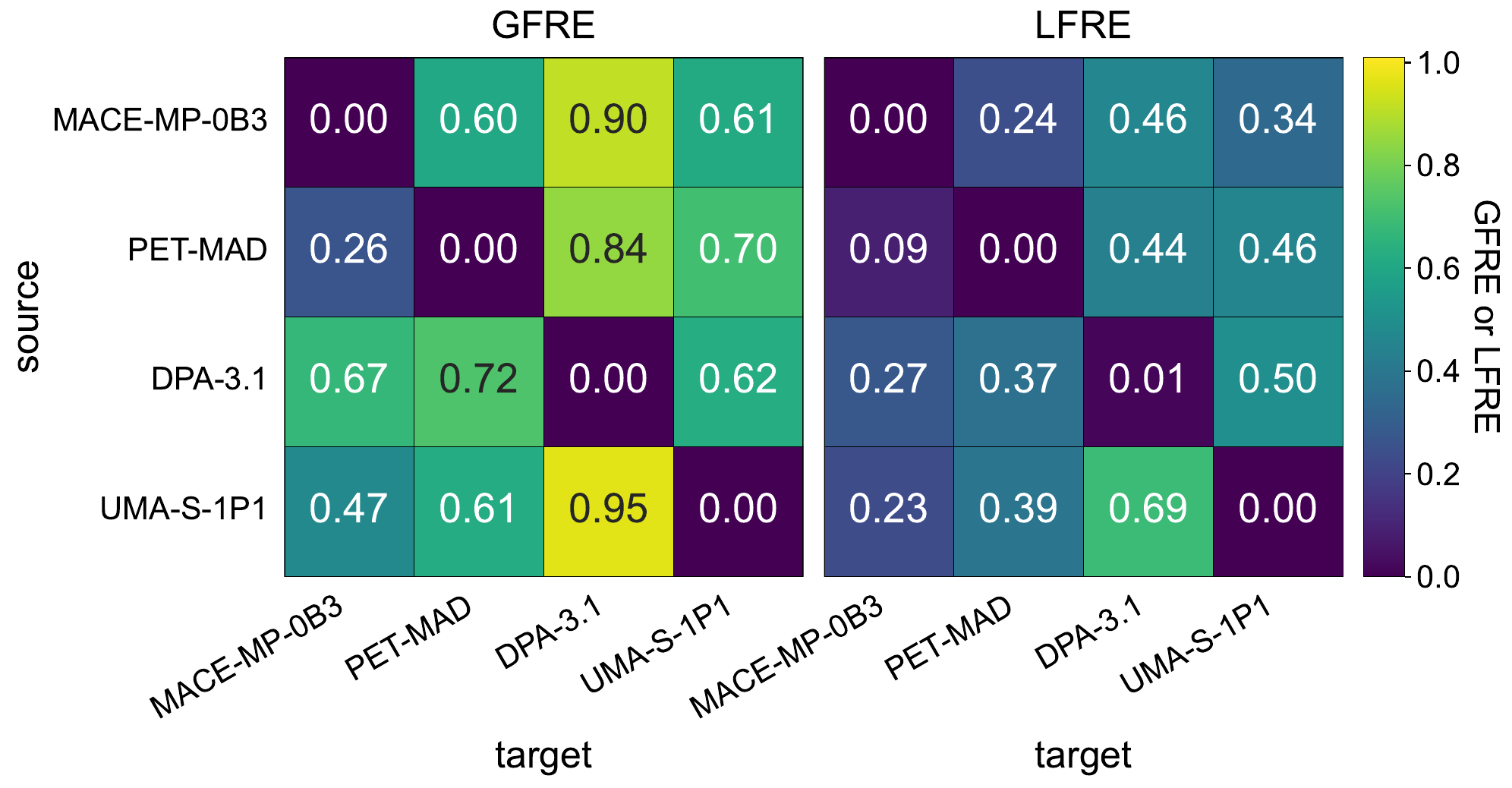}
    \caption{Heat maps of the global feature reconstruction error (GFRE, left) and local feature reconstruction error (LFRE, right) for the atomic last-layer latent features of MACE-MP-0b3, PET-MAD, DPA-3.1, and UMA-S-1P1, computed for the atomic environments from the test subset of the MAD dataset. Each cell represents the reconstruction error when mapping latent features of the ``source'' (row) to ``target'' (column).
    }
    \label{fig:umlip-errors}
\end{figure}

We first quantitatively compare the latent features of different uMLIPs by computing the GFRE and LFRE between the four models considered.
The errors are computed for structures in the test subset of the Massive Atomic Diversity (MAD) dataset~\cite{MAD}, which contains chemical systems with a broad coverage across 85 elements and exhibits vast structural diversity. We exclude the single-atom structures (incompatible with UMA) and polonium- or radon-containing structures (not supported by MACE trained on MPtrj) to yield 258,561 atomic environments from 9,461 structures. We then perform random stratified sampling across the subsets of MAD dataset at the atomic level to select 1,000 atomic environments for $\mathcal{D}_{\text{test}}$. All remaining environments are used as $\mathcal{D}_{\text{train}}$ to compute the regression weights $\widehat{P}_{FF'}$.

The heat maps in Fig.~\ref{fig:umlip-errors} reveal distinct trends in the feature reconstruction performances across the uMLIPs. Overall, the values are large (average of 0.66 for GFRE and 0.37 for LFRE off-diagonal cross-terms), indicating that each model encodes the chemical space in a unique manner. That latent spaces of different models differ significantly is not surprising. Latent features are only weakly constrained by the learning target: in principle, a model could learn the energy as the first feature, and have all others reflect completely unrelated components of the molecular structure. Therefore, the latent space reflects the priors explicitly and implicitly encoded in the model architecture, the choice of hyperparameters, and the training strategy.

The LFREs are generally lower than the GFREs, which reflects the ability of the LFRE to implicitly account for nonlinear relationships between the feature spaces. The lowest GFREs and LFREs are observed when MACE-MP-0b3 features are used as targets, suggesting that the other uMLIPs can reliably recover the information encoded in its features. Conversely, the highest errors are observed when DPA-3.1 features serve as targets.
The average GFREs across the off-diagonal targets for each source model are 0.70 for MACE-MP-0b3, 0.60 for PET-MAD, 0.67 for DPA-3.1, and 0.68 for UMA-S-1P1. For the LFRE, they are 0.35, 0.33, 0.38, and 0.44, respectively. PET-MAD exhibits marginally lower GFREs and LFREs in reconstructing the features of all other uMLIPs, despite being trained on the smallest dataset among the models considered. To rule out the fact that the low PET-MAD reconstruction errors are lower only because we are using the MAD dataset as a benchmark, we repeat the analysis using the Alexandria subset \cite{Alexandria} (cf. SI~\ref{sec:si-errors-umlip-alexandria}). We observe similar trends, indicating that the smaller MAD dataset can still sufficiently represent the configuration space of much larger datasets, and that its choice for our analysis does not introduce a significant bias.

In SI \ref{sec:si-errors-umlip-bb}, we also present the feature reconstruction errors (FREs) across the models for the backbone features (representations obtained immediately after message-passing and before the MLP readout), which show smaller reconstruction error (average off-diagonal GFRE is 0.54, LFRE is 0.30) and hence relatively higher generalizability across the models.

\subsection{Comparing model variants}
\label{sec:variants}

Beyond the comparison between entirely different uMLIPs, we assess the consistency of information content across uMLIP ``variants'' of the same architecture. Through this analysis, we establish the sensitivity of various architectures to changes in the training dataset or target. We also establish a scale for the ``internal'' variability of each architecture that helps calibrate the comparison between models that differ more substantially. We categorize the variants into: ``single-task'' models where an entirely separate model architecture is trained from scratch for distinct datasets, and ``multi-task'' models that either adopt a multi-head \cite{Caruana1997, ruder2017} or mixture-of-experts \cite{jacobs1991adaptive, riquelme2021scaling, shazeer2017outrageously} approach to simultaneously target multiple datasets with a single model architecture, often with significant weight sharing between the prediction tasks.

\begin{figure}[htbp]
    \centering
    \includegraphics[width=\linewidth]{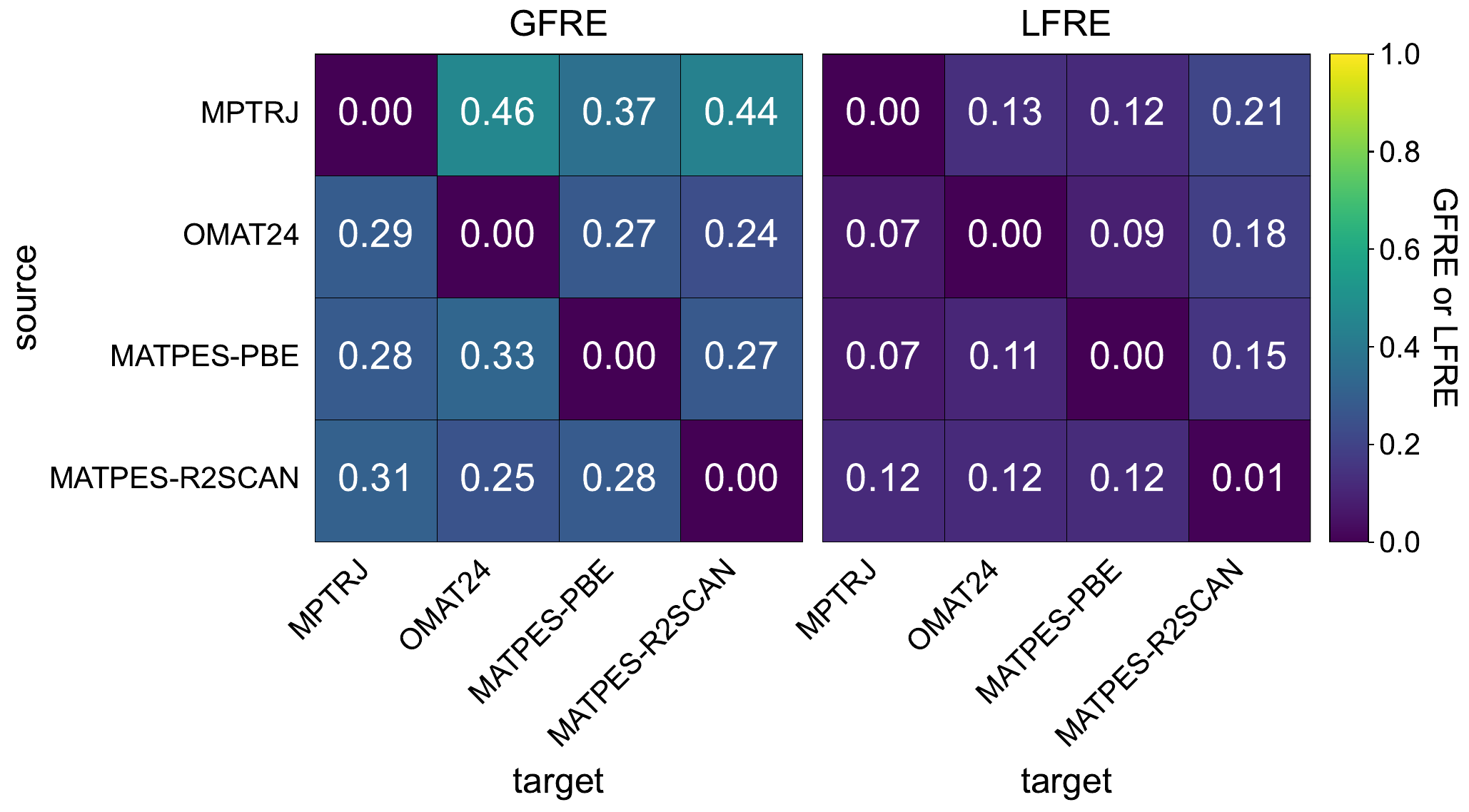}
    \caption{The reconstruction errors across single-task MACE models trained on MPtrj (MACE-MP-0b3), OMat24, and MatPES (PBE and r2SCAN) datasets~\cite{MPTRAJ, MATPES, omat24} evaluated on the MAD test subset.}
    \label{fig:variants-errors_mace}
\end{figure}

We first compare the reconstruction errors across the single-task MACE variants, which have been trained separately on MPtrj \cite{MPTRAJ}, OMat24 \cite{omat24}, MatPES-PBE, and MatPES-r2SCAN \cite{MATPES}. The latter two datasets only differ in the exchange-correlation functional used during the reference calculations, which are PBE~\cite{perd+96prl} and r2SCAN~\cite{Furness2020}, respectively. Overall, the reconstruction errors are moderately low across the variants (Fig.~\ref{fig:variants-errors_mace}) (the average off-diagonal GFRE is 0.31, and the average LFRE is 0.12).
The MPtrj variant exhibits the largest overall GFREs and LFREs (average GFRE and LFRE are 0.42 and 0.15, respectively), which reflects the limited information content of the smallest and oldest MPtrj dataset.  
The reconstruction errors between MatPES-PBE and MatPES-r2SCAN are comparable to the other cases, revealing that having different functionals in the reference calculations converges the resulting variant models into significantly distinct latent spaces. The lowest overall errors are observed for the OMat24 variant (on average, the GFRE is 0.25 and the LFRE is 0.09), which is indicative of the relatively rich information content of OMat24 dataset that is often utilized in the pre-training of uMLIPs \cite{eSEN, GRACE, mace_multihead, ORB3, omat24}.

In DPA-3.1, a model trained on the OpenLAM-v1 collection of 31 datasets \cite{OpenLAMv1}, the following multi-task learning strategy is adopted: the model incorporates a shared backbone network while still being conditioned for different datasets via one-hot encoding. This backbone is followed by multiple specialized output heads, or branches, tailored to each dataset \cite{DPA3}. Reconstruction results between branches corresponding to OMat24, MPtrj, OC20, ODAC, and SPICE datasets \cite{omat24, MPTRAJ, OC20, ODAC, SPICE}, presented in SI~\ref{sec:si-variants-dpa}, indicate comparable latent space consistency to the MACE single-task models, suggesting that strong parameter sharing without dynamic routing preserves highly universal latent representations.

\begin{figure}[htbp]
    \centering
    \includegraphics[width=\linewidth]{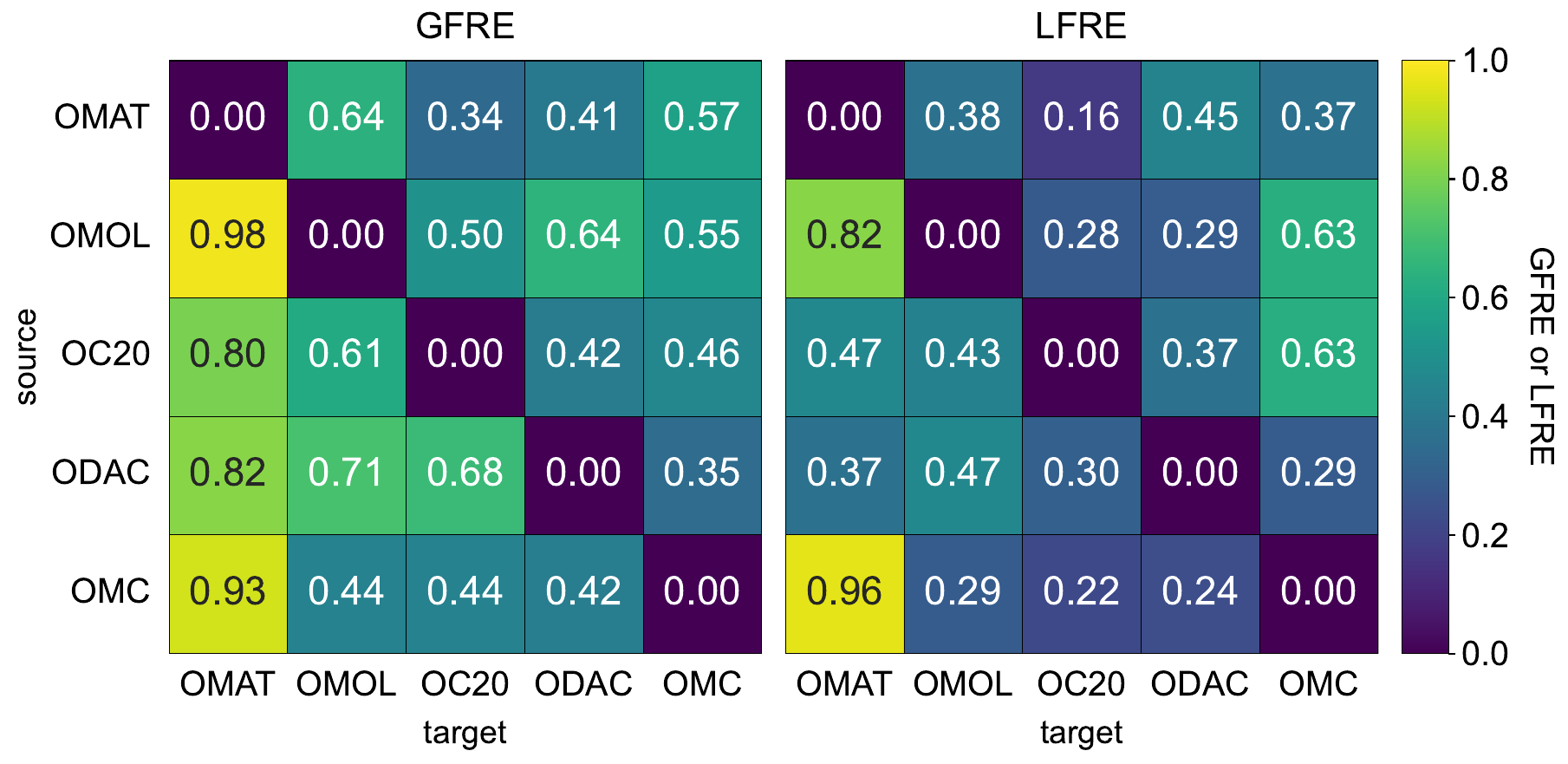}
    \caption{The reconstruction errors across different input tasks of UMA-S-1P1 trained on OMat24, OMOL, OC20, ODAC, and OMC \cite{omat24, OMOL, OC20, ODAC, OMC25} evaluated on the MAD test subset.
    }
    \label{fig:variants-uma_errors}
\end{figure}

UMA is a uMLIP developed with the goal of having a single model that performs well across all atomistic modeling tasks, encompassing materials, molecules, and their interactions. To do so, the model has been trained simultaneously on OMol25, OMat24, OC20, ODAC25, and OMC25 \cite{OMOL, omat24, OC20, ODAC, OMC25}. Instead of adopting a multi-head approach for the different datasets, the model incorporates a MoLE strategy in its interaction layers, where the weights of the different experts are computed based on the global embedding that encodes the dataset-specific information \cite{UMA}. UMA-S-1P1 shows greater variability in reconstruction errors (see Fig.~\ref{fig:variants-uma_errors}), which suggests that the MoLE mechanism encourages stronger specialization of feature manifolds corresponding to each dataset. We observe that the OMat24 task reconstructs other variants with relatively low errors (average off-diagonal GFRE is 0.49 and LFRE is 0.34), similar to previous results with MACE, whereas other variants struggle to approximate its latent space linearly. We also observe group-level consistency, as the features of catalysis-related tasks (OC20, ODAC25, OMC25) generally exhibit lower reconstruction errors (GFRE is 0.46, and LFRE is 0.34 on average across this group).

To investigate the influence of the training domain on the learned latent space, we also compare the organics-focused MACE-OFF23 model \cite{MACEOFF} with the materials-focused  MACE-MP-0b3 \cite{MACE} by evaluating the reconstruction errors on the organic structures of the MAD test subset. As detailed in SI~\ref{sec:si-variants-domains} (Table \ref{tab:mace-domains}), the materials-trained model reconstructs the organic-trained model almost perfectly  (GFRE is 0.02), whereas the reverse direction yields remarkably higher global error (GFRE is 0.36), despite similar local agreement (LFRE is 0.04). This could suggest that the broader materials manifold learned by MACE-MP-0b3 cannot be fully captured by MACE-OFF23, which has been exposed only to organic systems.

We further examine the effect of model capacity by comparing the MACE checkpoints of varying sizes (small, medium, large) within the same architecture family, for both the materials-focused MACE-MP-0a series and the organic-focused MACE-OFF23 series (SI~\ref{sec:si-variants-sizes}, Fig.~\ref{fig:mace-sizes}). Interestingly, regardless of size, all MACE-MP-0a variants reconstruct each other with near-perfect linear mapping, whereas reconstructing the large MACE-OFF23 checkpoint from its smaller counterparts is significantly harder (average GFRE is 0.48 when small/medium $\rightarrow$ large). A similar size-scaling analysis on PET-OMAT models \cite{uPET} (SI Fig.~\ref{fig:pet-omat}) shows that smaller models can still approximate the principal manifold of larger ones reasonably well, with the medium-sized variant offering a good information content/size tradeoff.

Finally, we assessed the impact of single-task versus multi-head readout architectures using two MACE models trained on the same OMat24 dataset: the original single-head MACE-OMAT-0 and the OMat-specific head of the multi-head MACE-MH-1 model \cite{mace_multihead} (the latter having undergone multi-head replay fine-tuning after initial pre-training). Results are presented in SI~\ref{sec:si-variants-mace-archictures} (Table~\ref{tab:variants-mace-architectures}). Both directions exhibit moderate global reconstruction error (GFRE is $\sim$ 0.33) but low local error (LFRE is less than 0.10), demonstrating that the multi-head replay protocol preserves the core local feature neighborhoods of the original single-head representations while allowing the readout to become more expressive.

\subsection{Varying targets}

Although we have primarily focused on uMLIPs, universal models are also available for properties other than the ground state potential energy surface. To this end, we compare the latent features of PET-MAD and PET-MAD-DOS \cite{PET-MAD-DOS-2025}. The two models have nearly identical architectures, differing only in the last linear readout layer due to the different sizes of the targets (last-layer features are still of the same dimension), and are both trained on the MAD dataset. While PET-MAD was trained on energies and forces, PET-MAD-DOS was trained on the electronic density of states (DOS). 

The results in SI~\ref{sec:si-dos} reveal that PET-MAD-DOS exhibits more information content than PET-MAD, as seen by the lower GFRE and LFRE values (0.68 and 0.39 for energy $\to$ DOS vs. 0.56 and 0.28 for DOS $\to$ energy). While this observation is expected, provided that DOS contains far more information than the energies and forces, it is still interesting to see that the information content in the latent features also reflects our physical knowledge of the targets. Still, the substantial reconstruction error in the DOS $\to$ energy direction indicates that the energy-based latent features also encode distinct information.

\subsection{Fine-tuning}
\label{sec:fine-tuning}

The advent of uMLIPs has made fine-tuning on domain-specific datasets a standard approach for obtaining accurate potentials tailored to specific applications that can still generalize well \cite{deng2024, Jacobs2025, Radova2025, Kaur2025}. Here, we analyze the reconstruction errors between features that arise from different fine-tuning strategies, as well as a model trained entirely from scratch. We also study the evolution of learned latent features along a ``training trajectory'' by computing the reconstruction errors of latent features sampled from model checkpoints at intermediate training epochs with respect to the fully trained model.

\begin{figure}[htbp]
    \centering
    \includegraphics[width=\linewidth]{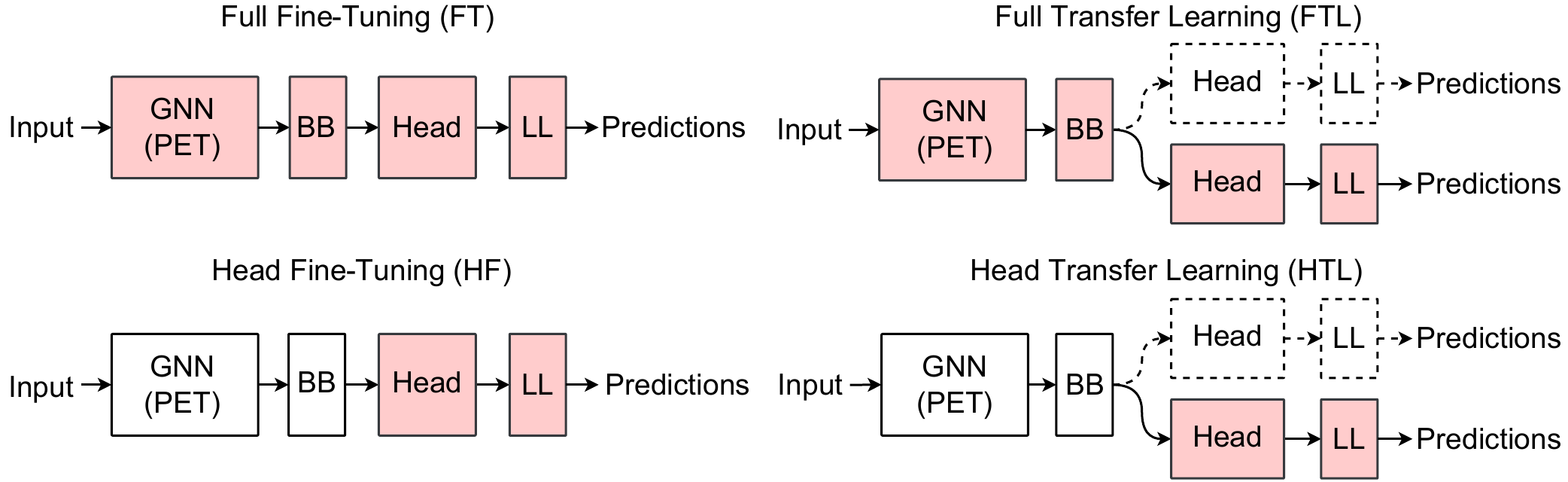}
    \caption{A schematic overview of the fine-tuning strategies for the PET architecture: full fine-tuning, head fine-tuning, full transfer learning, and head transfer learning. The graph neural network (GNN) based on the Point Edge Transformer (PET) computes the backbone (BB) features. For a given prediction target, these BB features are fed into the readout ``heads'' (multi-layer perceptrons or MLPs) to generate the last-layer (LL) features. The four illustrated strategies involve training different combinations of these model components, and the trainable parts are colored in salmon.}
    \label{fig:pet-fine-tuning}
\end{figure}

For our analysis, we consider the PET architecture and its pre-trained PET-MAD uMLIP as the baseline. We train material-specific MLIPs for lithium thiophosphate (LPS), a promising class of solid-state electrolytes widely studied for the next generation of battery materials \cite{liuAnomalousHighIonic2013a, fragapane2025lipselectrolytematerialsbenchmark,katoLithiumionconductiveSulfidePolymer2021,manthiramLithiumBatteryChemistries2017,tisiThermalConductivityLi2024,gigliMechanismChargeTransport2024}. We consider the following training or fine-tuning procedures: a bespoke model trained from scratch on the dataset from randomly initialized weights; full fine-tuning (FF), where all trainable model weights are optimized on the new dataset; ``head-only'' fine-tuning (HF), where only the weights of the MLP head are optimized during training while the rest remains fixed; full transfer learning (FTL), where a new MLP head is added with random weight initialization, and then all trainable weights of the model are optimized; head transfer learning (HTL), where a new MLP head is created with random weights, and only the MLP head weights are optimized. A schematic overview of the trainable parts for these fine-tuning strategies is given in Fig.~\ref{fig:pet-fine-tuning}. The training procedures were carried out using \textit{metatrain} \cite{metatrain} (see SI~\ref{sec:si-fine-tuning} for the error metrics and hyperparameters of the fine-tuned models). We use the dataset from Ref.~\cite{tuerk2025}, which was previously used to model complex surface structural reconstructions on LPS. The dataset contains 4,088 structures, yielding 225,905 atomic environments in total. In particular, it contains the structures of bulk LPS in each of its phases ($\gamma$, $\beta$, $\alpha$, and amorphous) and the structures of the LPS surface with and without absorbed water molecules.

\begin{figure}[htbp]
    \centering
    \includegraphics[width=\linewidth]{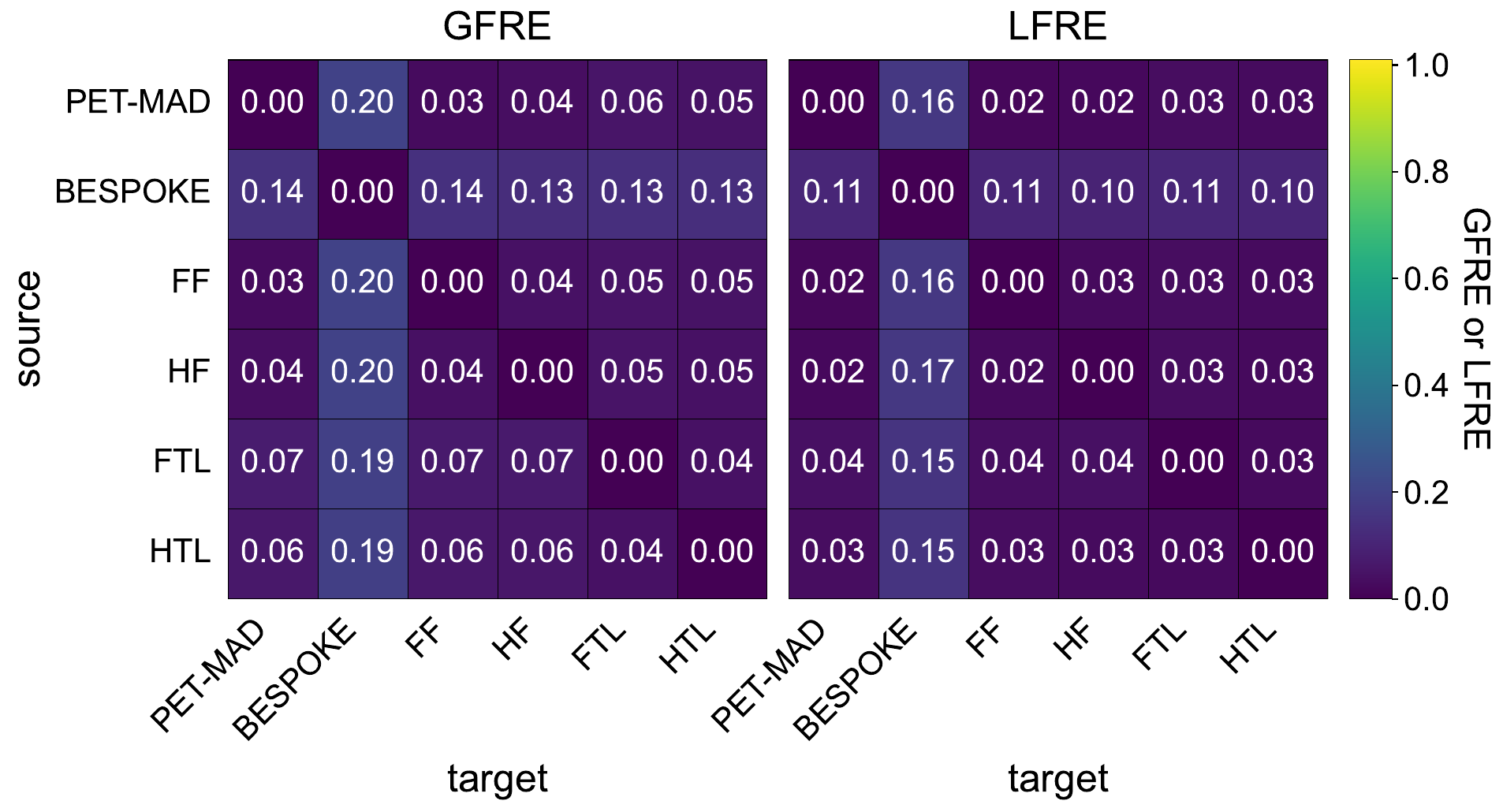}
    \caption{The reconstruction errors across the last-layer atom-level features of differently fine-tuned PET checkpoints computed for the LPS dataset.}
    \label{fig:lips-errors_pet}
\end{figure}

Fig.~\ref{fig:lips-errors_pet} shows the reconstruction errors among the final models from the described fine-tuning strategies, the pre-trained PET-MAD model, and a bespoke PET model trained from scratch on the LPS dataset.
The reconstruction errors are notably low between all of the fine-tuned models (FF, HF, FTL, and HTL) and the pre-trained PET-MAD model, which signifies a high degree of similarity in their last-layer features despite differences in training protocols. This suggests that the fine-tuned models retain a strong bias from the pre-trained model, which is also evident in the PCovR projections of the last-layer features (see Fig.~\ref{fig:lips-pcovr}).
The bespoke model exhibits the highest reconstruction errors, yet the values are still moderate, revealing that its training has converged to a distinct yet nearby minima with respect to the other models.

\begin{figure}[htbp]
    \centering
    \includegraphics[width=\linewidth]{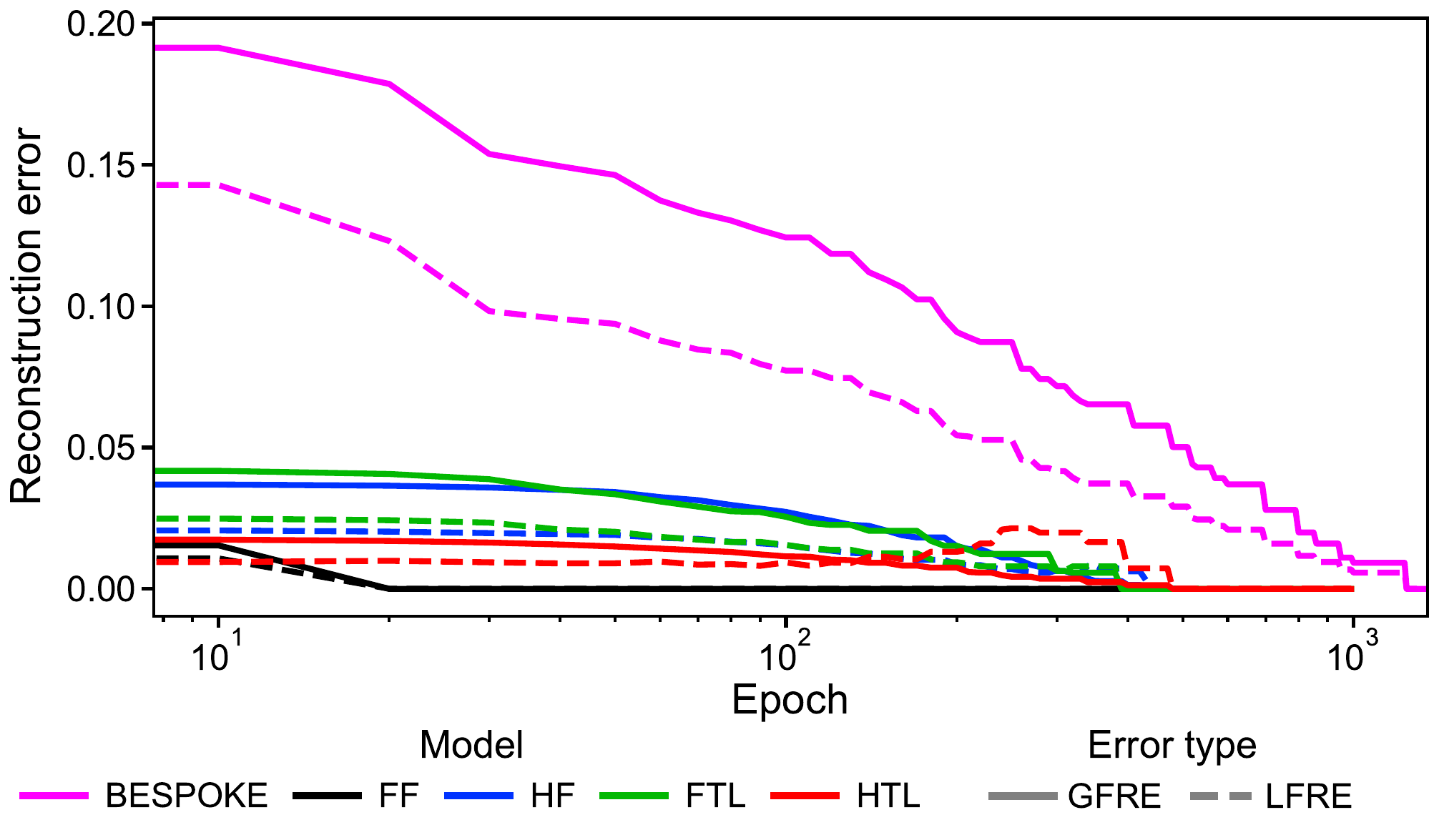}
    \caption{The evolution of the reconstruction errors of last-layer atom-level features during training on the LPS dataset, where the errors have been computed with respect to the features of fully trained models corresponding to each respective fine-tuning strategy as targets.
    }
    \label{fig:lips-evolution-erorrs-lips}
\end{figure}

Fig.~\ref{fig:lips-evolution-erorrs-lips} shows the evolution of the reconstruction errors along the training trajectory computed relative to the fully converged model for each strategy. The bespoke model trained from scratch starts with high GFRE and LFRE, which gradually decay, yet remain above numerical zero even after 1000 epochs (0.009 for GFRE and 0.005 for LFRE). In contrast, fine-tuned strategies begin with lower initial reconstruction errors due to pre-conditioning with the PET-MAD uMLIP, which then converge quickly to near-zero values. In FTL and HTL, the curve is not monotonic, and more than 400 epochs are needed for convergence, most likely due to the randomized head initialization. In HF, a similar number of epochs is needed to align the features with those of the final model, whereas FF converges the reconstruction errors to $< 0.01$ with only 20 epochs. This can be explained by the relative differences in the training degrees of freedom between the two strategies.
Overall, the combination of low reconstruction errors in Fig.~\ref{fig:lips-errors_pet} and rapid convergence in Fig.~\ref{fig:lips-evolution-erorrs-lips} demonstrates that the last-layer features of PET-MAD provide a robust starting point for the fine-tuning strategies. This provides further explanation as to why the uMLIPs can converge rapidly to high accuracy when fine-tuned to a new, narrower dataset \cite{Kaur2025}.

\subsection{Backbone vs. last-layer features}

While we have chosen the last-layer (LL) features for our analysis thus far, the latent features can be extracted from any other layer within the NN architectures of the uMLIPs. Another features of particular interest are the ``backbone'' (BB) latent features, which correspond to the representations obtained immediately after all message-passing iterations, but before entering a multilayer perceptron (MLP) head, which is an architectural choice followed by many uMLIPs (see Fig.~\ref{fig:pet-fine-tuning} for a sample schematic of PET-MAD). The BB features are of particular interest in multi-head and multi-task models, where they often serve as a shared input across the multiple heads. To provide examples of both multi-task and single-task contexts, we evaluate the reconstruction errors between the last-layer features and the backbone features for UMA and PET.

\begin{figure}[htbp]
    \centering
    \includegraphics[width=\linewidth]{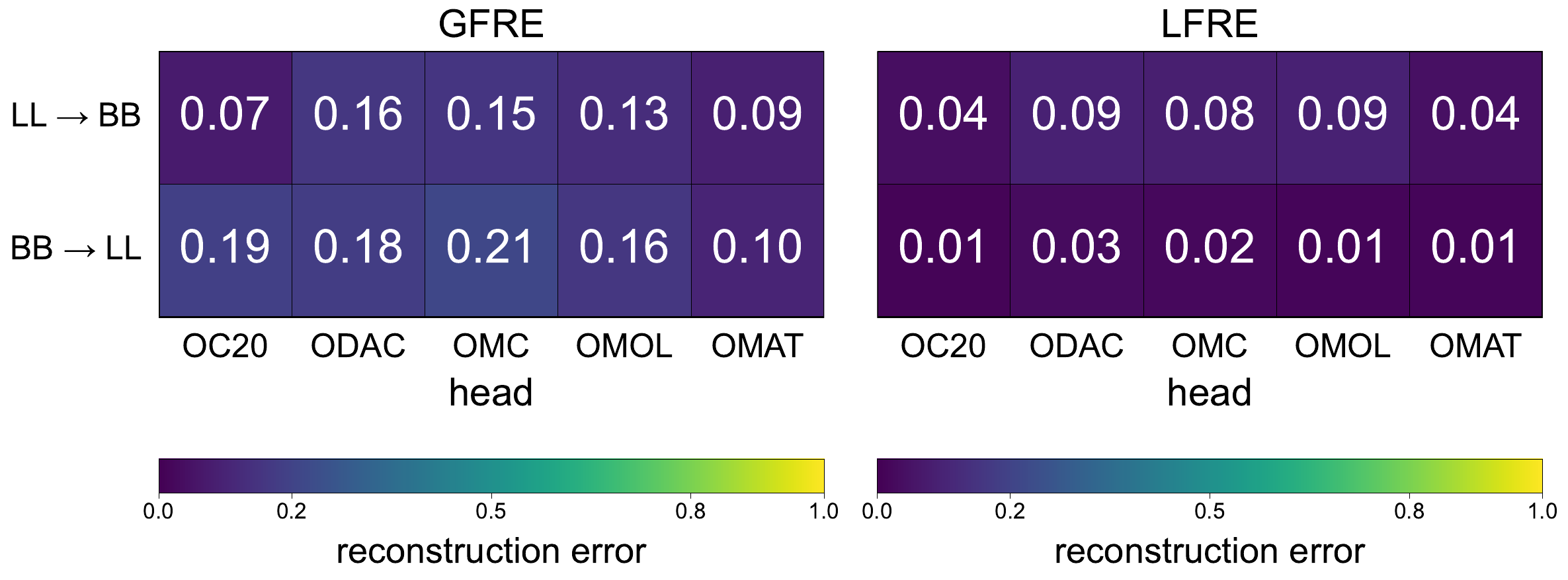}
    \caption{The global and local reconstruction errors between the last-layer features (LL) and backbone features (BB) of UMA-S-1P1 for different experts, computed for the MAD test subset.
    }
    \label{fig:ll_vs_bb-uma}
\end{figure}

In UMA-S-1P1 (Fig.~\ref{fig:ll_vs_bb-uma}), the BB $\to$ LL and LL $\to$ BB GFREs are moderate and similar in both directions, consistent with the fact that a global linear mapping cannot capture the nonlinear effects of the MLP. On the contrary, the LFREs are lower, as they can locally capture the said transformation. The LL $\rightarrow$ BB LFRE is larger than vice versa, suggesting that there exists some information loss in going from the shared BB features to the specialized LL features, or equivalently, that the shared features contain more information, making them suitable for specialization into different prediction tasks.

For PET (Fig.~\ref{fig:ll_vs_bb-pet} in SI~\ref{sec:si-ll-vs-bb}), both reconstruction errors are higher and more varied across model variants compared to UMA-S-1P1, indicating greater divergence between BB and LL representations. Even in the GFRE, reconstructing BB from LL yields higher errors than the reverse case.
Once again, the LFRE is lower in both reconstruction tasks, but the improvement is only minor in the case of LL $\rightarrow$ BB. We speculate that the large information loss might be associated with the MLP eliminating the non-invariant parts of the BB features, which takes place as the PET model learns the symmetries of the energy prediction task.

These asymmetries in the embedding reconstruction of the two uMLIPs highlight the dominant influence of model architecture on latent feature encodings, with design choices controlling the tradeoff between shared general-purpose representations and task-specific final refinements.

\subsection{Local to global features}
\label{sec:cumulants}

While most uMLIPs strictly work at the level of atom-centered local environments, one often needs to operate at the molecular or structural level, e.g., to build data-driven maps of datasets or to analyze the outcome of simulations \cite{trib+12pnas, isay+15cm}. For these applications, it is customary to average the atomic descriptors to obtain global features \cite{Guo2022, Cheng2020}, which, however, entirely disregards the variability across the atomic environments and results in significant information loss. To mitigate this and capture comprehensive structural information, we investigate the use of higher-order cumulants of atomic features to construct structure-level descriptors.

To formalize this, consider a dataset $\mathcal{D}$ consisting of structures, where each structure $S$ has $N_S$ atoms. For a given model $F$, the atomic feature matrix for a structure $S$ is
\begin{equation}
X_S^{F} =
\begin{bmatrix}
    \boldsymbol{\xi}_{S,1}^{F} \\
    \boldsymbol{\xi}_{S,2}^{F} \\
    \vdots \\
    \boldsymbol{\xi}_{S,N_S}^{F} \\
\end{bmatrix}
\in \mathbb{R}^{N_S \times d_F},
\end{equation}
where $\boldsymbol{\xi}_{S,i}^{F} \in \mathbb{R}^{d_F}$ is the feature vector of the atom $i$ in the structure.

We construct structure-level descriptors by concatenating the first $M$ cumulants of the atomic feature distribution. Let $\boldsymbol{\kappa}_S^{(k)}$ denote $k$-th order cumulant vector for structure $S$. These are obtained from the central moments of the atomic features. Specifically, the mean feature vector is:
\begin{equation}
\bar{\boldsymbol{\xi}}_S^{F} = \frac{1}{N_S}
\sum_{i=1}^{N_S} \boldsymbol{\xi}_{S, i}^{F},
\end{equation}

\noindent and for $k \geq 2$, the $k$-th central moment vector is:
\begin{equation}
\boldsymbol{\mu}^{(k)}_S = \frac{1}{N_S} \sum_{i=1}^{N_S} \left(
\boldsymbol{\xi}_{S,i}^{F} - \bar{\boldsymbol{\xi}}_S^{F}
\right)^k,
\end{equation}

\noindent where the power is applied entry-wise to each feature. The cumulants $\boldsymbol{\kappa}^{(k)}_S$ are then derived as follows (full formulae up to order 8 are provided in the SI~\ref{sec:si-cumulants}):

\begin{align}
\boldsymbol{\kappa}^{(1)}_S &= \bar{\boldsymbol{\xi}}_S^{F}, \\
\boldsymbol{\kappa}^{(2)}_S &= \boldsymbol{\mu}^{(2)}_S, \\
\boldsymbol{\kappa}^{(3)}_S &= \boldsymbol{\mu}^{(3)}_S, \\
\boldsymbol{\kappa}^{(4)}_S &= \boldsymbol{\mu}^{(4)}_S - 3 (\boldsymbol{\mu}^{(2)}_S)^2, \\
\boldsymbol{\kappa}^{(5)}_S &= \boldsymbol{\mu}^{(5)}_S - 10 \boldsymbol{\mu}^{(2)}_S \boldsymbol{\mu}^{(3)}_S.
\end{align}

For numerical stability and to keep the scale of the descriptors comparable across orders, we use signed root cumulants:

\begin{equation}
\tilde{\boldsymbol{\kappa}}_S^{(k)} = \mathrm{sign}(\boldsymbol{\kappa}_S^{(k)}) \cdot |\boldsymbol{\kappa}_S^{(k)}|^{1/k}.
\end{equation}

\noindent The cumulative structure-level descriptor up to order $M$ is then a concatenation:
\begin{equation}
\phi_S^{(M)} = \left[ \tilde{\boldsymbol{\kappa}}_S^{(1)}, \tilde{\boldsymbol{\kappa}}_S^{(2)}, \dots, \tilde{\boldsymbol{\kappa}}_S^{(M)} \right].
\end{equation}

\noindent To determine the order at which these descriptors converge, i.e., when additional moments add negligible new statistics, we compute the reconstruction errors using the last-layer atomic features of PET-MAD on the MAD test set.

\begin{figure}[htbp]
    \centering
    \includegraphics[width=\linewidth]{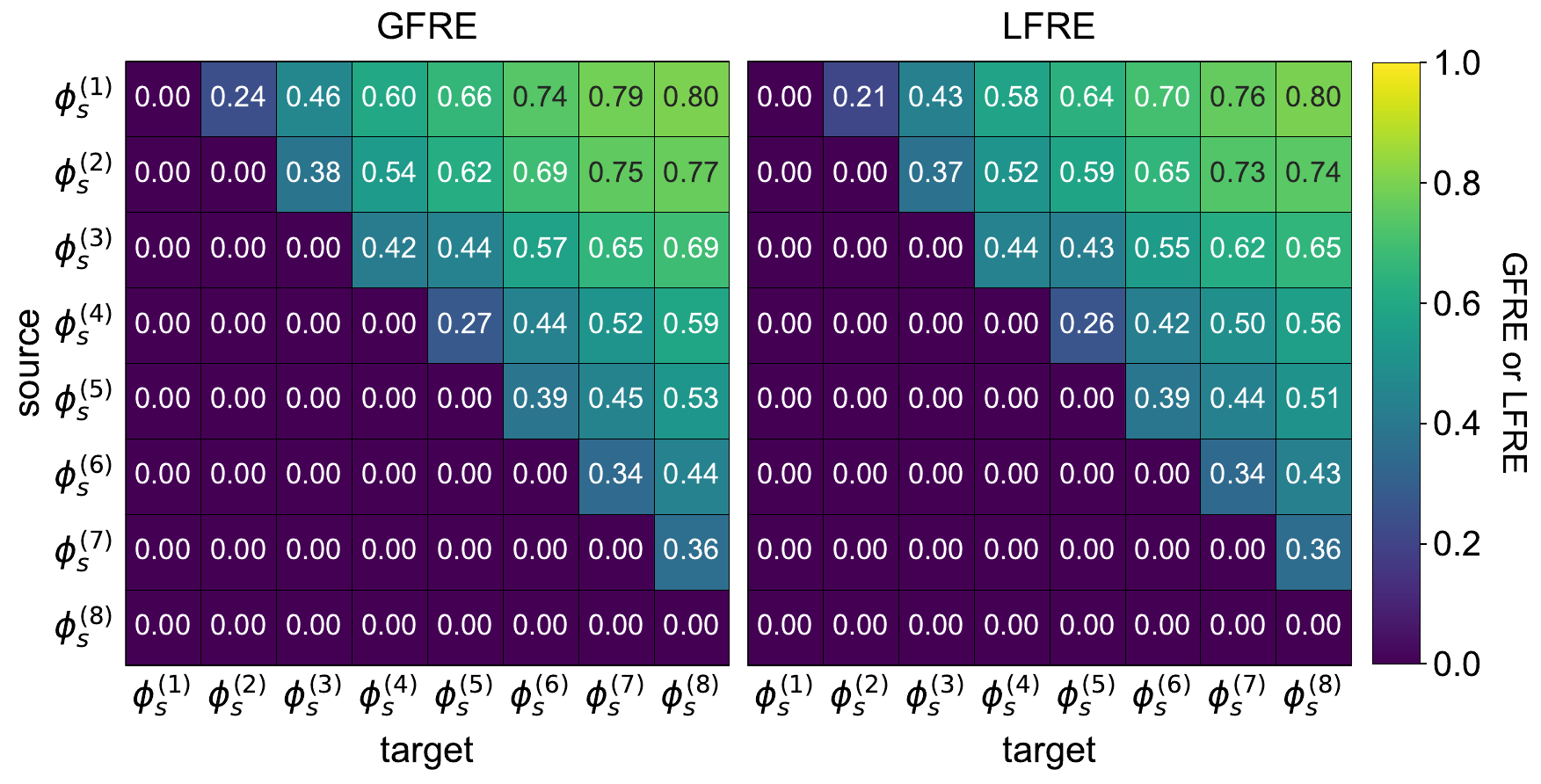}
    \caption{The reconstruction errors of the progressive cumulative structure-level features, constructed as concatenations of central moments of last-layer features per atom. The features are derived from PET-MAD for the MAD test subset. $\phi_S^{(m)}$ denote cumulative descriptors up to order $m$. For the local feature reconstruction error (LFRE), we set the neighborhood size $k$ to the larger feature dimension between the compared feature pairs.}
    \label{fig:moments-moments_errors}
\end{figure}

The reconstruction errors for cumulative structure-level descriptors (Fig.~\ref{fig:moments-moments_errors}) show a hierarchical information flow, where higher-order moments fully sublimate lower-order ones. In contrast, lower-order descriptors cannot reconstruct higher ones, with errors scaling with order. This asymmetry is expected as the higher moments encode novel distributional aspects, such as skewness, beyond the mean and variance. 
This implies that structures with limited variability in the atomic environments, e.g., pristine crystals, lower orders may suffice in the global feature construction, but for most systems with high local variability, higher moments are needed to capture rare or asymmetric contributions. While the errors decrease progressively, they do not yet saturate by the eighth order: GFRE and LFRE for reconstructing the eighth order from the seventh are still above 0.3. This hints that more statistically meaningful structural inhomogeneities would be captured by going to even higher cumulants.

Applying eighth-order cumulative descriptors to cross-model analysis on the same MAD test dataset (SI~\ref{sec:si-cumulants} Fig.~\ref{fig:cumulants-errors-umlip}) yields patterns similar to atomic-level features (Fig.~\ref{fig:umlip-errors}) but with amplified dissimilarities. Overall, the errors are much higher (on average, the GFRE is now 0.87, and LFRE is 0.83). This pattern suggests that the differences in the model latent features arising from the rare or asymmetric atomic-environment contributions is better captured by the higher cumulants. Practically, if the goal is to discriminate subtle model embeddings and emphasize rare motifs, inclusion of the higher cumulants is highly recommended.

\begin{figure}[htbp]
    \centering
    \includegraphics[width=\linewidth]{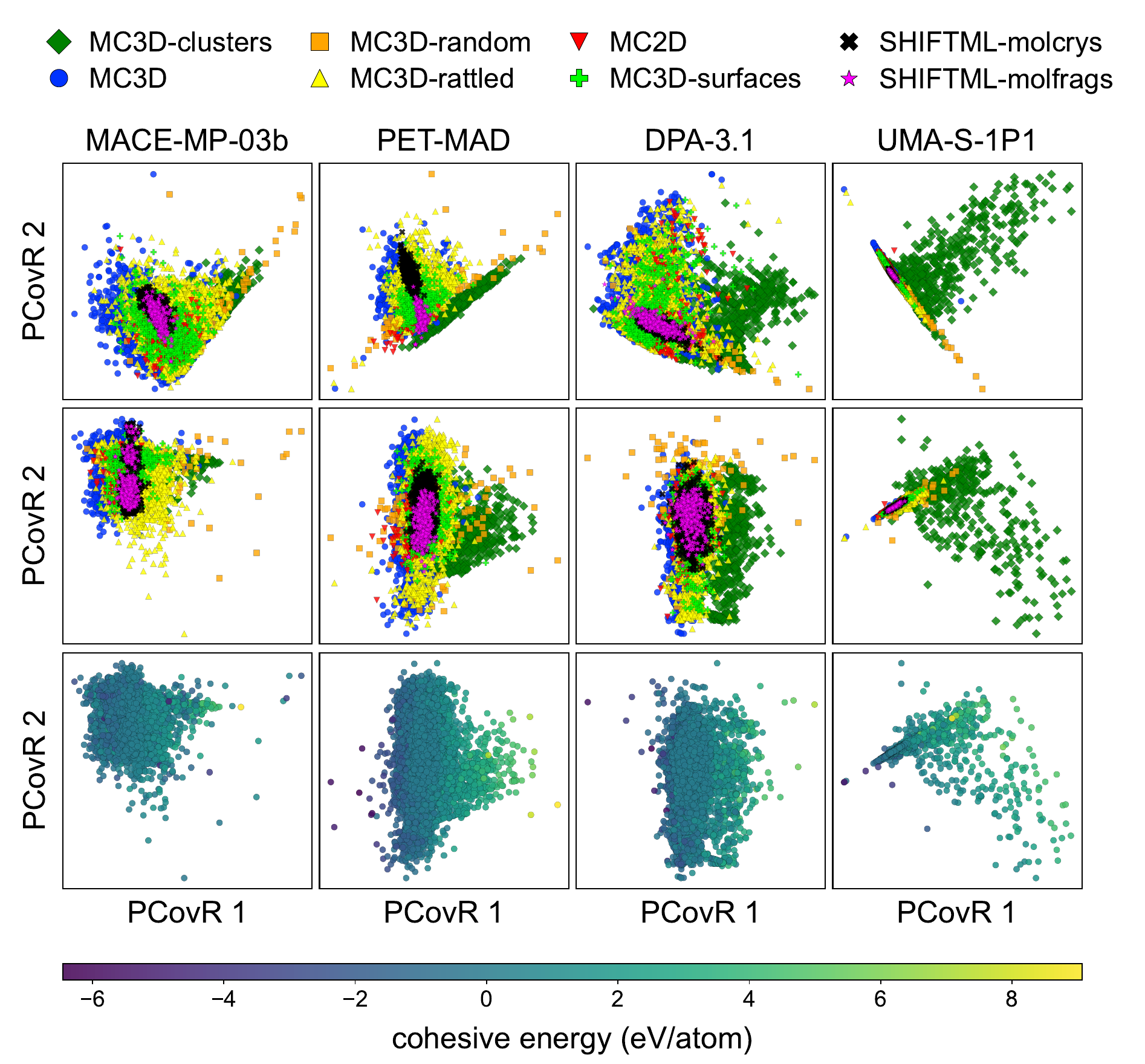}
    \caption{The PCovR projections of structure-level descriptors for different uMLIPs on the MAD test subset. Top row: projections using mean atomic features ($\tilde{\boldsymbol{\kappa}}_S^{(1)}$), colored by the different structural subsets of the MAD dataset. Middle row: projections using eighth-order cumulative cumulant features ($\phi_S^{(8)}$), again colored by respective structural subsets. Bottom row: same eighth-order cumulant features colored by the structural cohesive energy used as target in PCovR analysis.}
    \label{fig:cumulants-umlip-pcovr}
\end{figure}

To further illustrate the impact of higher-order statistics, we compare PCovR projections of structure-level descriptors built from mean-only features ($\tilde{\boldsymbol{\kappa}}_S^{(1)}$) with those using full eighth-order cumulative cumulants ($\phi_S^{(8)}$) (Fig.~\ref{fig:cumulants-umlip-pcovr}). When only the mean atomic features are used (top row), the projections exhibit a loosely consistent motif, characterized by a linear cluster corresponding to a few of the subsets and then the remaining subsets dispersed around it. Even at this level, however, the exact shape and subset assignment around the projection motif vary across the different models. For the eighth-order cumulant descriptors (middle row), the linear cluster fragments disappear entirely for MACE-MP-0b3, PET-MAD, and DPA-3.1, and the overall point distributions become highly model-specific with almost no cross-model alignment. Such vast differences are consistent with the higher values of the GFRE and LFRE observed between the uMLIPs in SI \ref{sec:si-cumulants} Fig.~\ref{fig:cumulants-errors-umlip}.

\section{Conclusions}

The rapidly growing number of ``universal'' atomistic machine learning potentials and models, differing in architecture and training strategy but capable of achieving similar accuracies on benchmarks, raises questions of how the different models achieve their learning targets, especially in terms of their internal representations of the chemical systems. 
In this work, we have carried out systematic comparisons of the latent spaces of several universal ML models and their variants, using the feature reconstruction error metrics that quantitatively compare the ``information content'' of different sets of features. 
We see that the reconstruction errors are generally high between the latent features of different uMLIPs, revealing the presence of significant differences between the internal representations of the models despite their comparable accuracy.
The local feature reconstruction errors are lower, which indicate the existence of nonlinear relations between the features of different models. They are nevertheless substantial, which suggests that different models choose to retain distinct structural and chemical information.

When comparing the variants of the same model architecture trained on different ``universal'' datasets, the trend becomes highly dependent on the adopted training strategy, with single-task and multi-head models exhibiting higher similarities than a model trained with the MoLE.
We also note that the model variant corresponding to the large OMat24 dataset generally exhibits the smallest reconstruction errors compared to the other cases.
This hints that a large, diverse dataset is not only beneficial for pre-training, but also encourages the learning of richer internal representations.
A similar trend is observed when comparing the PET-MAD with PET-MAD-DOS, a model built from the same architecture and training set but targeting the energy-resolved electronic density of states, thereby yielding more information-rich features.

The role of training becomes apparent when considering the analysis of models fine-tuned on a more specific dataset. We observe that all fine-tuning strategies allow near-perfect reconstruction from and to the latent space of the original model. A model trained entirely from scratch on the small, domain-specific dataset achieves moderate (but larger than for fine-tuned models) reconstruction errors against those fine-tuned from a uMLIP, at least within the narrow portion of configuration space covered by the specific dataset. 

The comparison between the backbone and last-layer features reveals that the backbone layer contains relatively richer information, which retrospectively explains the success of using multiple heads for different prediction targets. 
In the particular case of PET-MAD, a rotationally unconstrained model, we further attribute the large information asymmetry to the fact that the deeper layers of PET would exhibit a non-invariant character that gets successively projected out as the model approaches the prediction of the scalar energy target.
Lastly, we have extended the raw, atomic, local latent features of the uMLIPs to structural, global descriptors via higher-order cumulants.
The higher cumulants fully subsume the lower-order statistics, which critically indicate that significant information becomes lost when using the averages of atomic features as structural descriptors, which is a common practice in the field.

Our results yield several practical insights for atomistic ML practitioners. In particular, one can apply the feature reconstruction error metrics to understand the information flow within a model architecture, for example, by quantifying how much generality is lost in the readout layers. One could also imagine using the GFRE and LFRE to guide the optimization of architectural hyperparameters to maximize the descriptive power of internal representations.
In the context of fine-tuning, the changes in the latent features measured via the FREs can be linked to catastrophic forgetting and used to monitor the adjustments of the internal representations and prevent any significant loss of generalizability to out-of-distribution configurations. Given the hidden diversity of the uMLIP latent spaces uncovered in our study, we argue that predictive accuracy alone cannot sufficiently characterize atomistic ML models. Through the adoption of feature reconstruction error metrics, we establish a principled foundation for a more transparent, interpretable, and robust design of future atomistic ML models.

\section*{Software and Data}
The code and data required to reproduce the results of this work are available as a record \cite{matcloud_record} within the Materials Cloud \cite{Talirz_2020} Archive (DOI: 10.24435/materialscloud:r5-vh). Model weights used in this work were obtained from their public repositories:   \texttt{mace-foundations} (\url{https://github.com/ACEsuit/mace-foundations}), PET-MAD (\url{https://github.com/lab-cosmo/pet-mad}), DPA-3.1-3M (\url{https://www.aissquare.com/models/detail?name=DPA-3.1-3M&id=343&pageType=models}), and \texttt{fairchem} (\url{https://fair-chem.github.io/}). Original datasets can be obtained from the respective cited sources.

\begin{acknowledgments}

We thank A. Mazitov and F. Bigi for helpful discussions about the use of PET-MAD and its fine-tuning strategies. SoC, SaC, CM, DT, and MC acknowledge support from NCCR-MARVEL, funded by the Swiss National Science Foundation (SNSF) (grant number 205602). SaC and MC acknowledge support from a SNSF grant (project ID 200020\_214879). DT and MC acknowledge support from a Sinergia grant of the SNSF (grant ID CRSII5\_202296). WBH and MC acknowledges support from the European Research Council (ERC) under the research and innovation program (Grant Agreement No. 101001890-FIAMMA). This work was supported by grants from the Swiss National Supercomputing Centre (CSCS) under the projects s1243, s1219, lp26, and lp95.

\end{acknowledgments}

\bibliography{references.bib}

@Article{	  isay+15cm,
  archiveprefix	= {arXiv},
  arxivid	= {arXiv:1412.4096v3},
  author	= {Isayev, Olexandr and Fourches, Denis and Muratov, Eugene
		  N. and Oses, Corey and Rasch, Kevin and Tropsha, Alexander
		  and Curtarolo, Stefano},
  journal	= {Chemistry of Materials},
  pages		= {735--743},
  title		= {{Materials cartography: Representing and mining materials
		  space using structural and electronic fingerprints}},
  volume	= {27},
  year		= {2015}
}

@Article{	  perd+96prl,
  author	= {Perdew, Jp P and Burke, K and Ernzerhof, M},
  chapter	= {3865},
  journal	= {Phys. Rev. Lett.},
  pages		= {3865},
  publisher	= {American Physical Society},
  series	= {Phys. Rev. Lett. (USA)},
  title		= {{Generalized Gradient Approximation made simple}},
  volume	= {77},
  year		= {1996}
}

@Article{	  de+16jci,
  archiveprefix	= {arXiv},
  arxivid	= {1611.06246},
  author	= {De, Sandip and Musil, Felix and Ingram, Teresa and
		  Baldauf, Carsten and Ceriotti, Michele},
  journal	= {Journal of Cheminformatics},
  pages		= {6},
  publisher	= {Springer International Publishing},
  title		= {{Mapping and classifying molecules from a high-throughput
		  structural database}},
  volume	= {9},
  year		= {2017}
}

@Article{	  trib+12pnas,
  author	= {Tribello, Gareth A and Ceriotti, Michele and Parrinello,
		  Michele},
  journal	= {Proc. Natl. Acad. Sci. USA},
  pages		= {5196--201},
  pmid		= {22427357},
  title		= {{Using sketch-map coordinates to analyze and bias
		  molecular dynamics simulations.}},
  volume	= {109},
  year		= {2012}
}

@Article{	  tuerk2025,
  title		= {Reconstructions and Dynamics of
		  $\ensuremath{\beta}$-Lithium Thiophosphate Surfaces},
  author	= {T\"urk, Hanna and Tisi, Davide and Ceriotti, Michele},
  journal	= {PRX Energy},
  volume	= {4},
  issue		= {3},
  pages		= {033010},
  numpages	= {13},
  year		= {2025},
  month		= {Aug},
  publisher	= {American Physical Society},
  doi		= {10.1103/5hf9-hlj6},
  url		= {https://link.aps.org/doi/10.1103/5hf9-hlj6}
}

@Article{	  liuanomaloushighionic2013a,
  title		= {Anomalous {{High Ionic Conductivity}} of {{Nanoporous}}
		  Beta-{{Li}}{\textsubscript{3}} {{PS}}{\textsubscript{4}}},
  author	= {Liu, Zengcai and Fu, Wujun and Payzant, E. Andrew and Yu,
		  Xiang and Wu, Zili and Dudney, Nancy J. and Kiggans, Jim
		  and Hong, Kunlun and Rondinone, Adam J. and Liang,
		  Chengdu},
  year		= {2013},
  month		= jan,
  journal	= {J. Am. Chem. Soc.},
  volume	= {135},
  number	= {3},
  pages		= {975--978},
  issn		= {0002-7863, 1520-5126},
  doi		= {10.1021/ja3110895},
  urldate	= {2025-01-28},
  langid	= {english}
}

@Article{	  katolithiumionconductivesulfidepolymer2021,
  title		= {Lithium-Ion-Conductive Sulfide Polymer Electrolyte with
		  Disulfide Bond-Linked {{PS4}} Tetrahedra for
		  All-Solid-State Batteries},
  author	= {Kato, Atsutaka and Yamamoto, Mari and Utsuno, Futoshi and
		  Higuchi, Hiroyuki and Takahashi, Masanari},
  year		= {2021},
  month		= nov,
  journal	= {Commun Mater},
  volume	= {2},
  number	= {1},
  pages		= {112},
  issn		= {2662-4443},
  doi		= {10.1038/s43246-021-00216-0},
  urldate	= {2024-11-25},
  langid	= {english}
}

@Article{	  manthiramlithiumbatterychemistries2017,
  title		= {Lithium Battery Chemistries Enabled by Solid-State
		  Electrolytes},
  author	= {Manthiram, Arumugam and Yu, Xingwen and Wang, Shaofei},
  year		= {2017},
  month		= feb,
  journal	= {Nat Rev Mater},
  volume	= {2},
  number	= {4},
  pages		= {16103},
  issn		= {2058-8437},
  doi		= {10.1038/natrevmats.2016.103},
  urldate	= {2024-11-25},
  langid	= {english}
}

@Article{	  tisithermalconductivityli2024,
  title		= {Thermal Conductivity of {{Li}} 3 {{PS}} 4 Solid
		  Electrolytes with {\emph{Ab Initio}} Accuracy},
  author	= {Tisi, Davide and Grasselli, Federico and Gigli, Lorenzo
		  and Ceriotti, Michele},
  year		= {2024},
  month		= jun,
  journal	= {Phys. Rev. Materials},
  volume	= {8},
  number	= {6},
  pages		= {065403},
  issn		= {2475-9953},
  doi		= {10.1103/PhysRevMaterials.8.065403},
  urldate	= {2024-11-25},
  langid	= {english}
}

@Article{	  giglimechanismchargetransport2024,
  title		= {Mechanism of {{Charge Transport}} in {{Lithium
		  Thiophosphate}}},
  author	= {Gigli, Lorenzo and Tisi, Davide and Grasselli, Federico
		  and Ceriotti, Michele},
  year		= {2024},
  month		= feb,
  journal	= {Chem. Mater.},
  volume	= {36},
  number	= {3},
  pages		= {1482--1496},
  issn		= {0897-4756, 1520-5002},
  doi		= {10.1021/acs.chemmater.3c02726},
  urldate	= {2024-02-23},
  langid	= {english}
}

@Misc{		  fragapane2025lipselectrolytematerialsbenchmark,
  title		= {Li-P-S Electrolyte Materials as a Benchmark for
		  Machine-Learned Interatomic Potentials},
  author	= {Natascia L. Fragapane and Volker L. Deringer},
  year		= {2025},
  eprint	= {2511.16569},
  archiveprefix	= {arXiv},
  primaryclass	= {cond-mat.mtrl-sci},
  url		= {https://arxiv.org/abs/2511.16569}
}

@Article{	  goscinski_2021,
  doi		= {10.1088/2632-2153/abdaf7},
  url		= {https://doi.org/10.1088/2632-2153/abdaf7},
  year		= {2021},
  month		= {apr},
  publisher	= {IOP Publishing},
  volume	= {2},
  number	= {2},
  pages		= {025028},
  author	= {Goscinski, Alexander and Fraux, Guillaume and Imbalzano,
		  Giulio and Ceriotti, Michele},
  title		= {The role of feature space in atomistic learning},
  journal	= {Machine Learning: Science and Technology}
}

@Article{	  mad,
  author	= {Mazitov, Arslan and Chorna, Sofiia and Fraux, Guillaume
		  and Bercx, Marnik and Pizzi, Giovanni and De, Sandip and
		  Ceriotti, Michele},
  date		= {2025/11/21},
  doi		= {10.1038/s41597-025-06109-y},
  id		= {Mazitov2025},
  isbn		= {2052-4463},
  journal	= {Scientific Data},
  number	= {1},
  pages		= {1857},
  title		= {Massive Atomic Diversity: a compact universal dataset for
		  atomistic machine learning},
  url		= {https://doi.org/10.1038/s41597-025-06109-y},
  volume	= {12},
  year		= {2025}
}

@Article{	  scikit-matter,
  title		= {scikit-matter: A Suite of Generalisable Machine Learning
		  Methods Born out of Chemistry and Materials Science},
  author	= {Goscinski, Alexander and Principe, Victor Paul and Fraux,
		  Guillaume and Kliavinek, Sergei and Helfrecht, Benjamin
		  Aaron and Loche, Philip and Ceriotti, Michele and
		  Cersonsky, Rose Kathleen},
  journal	= {Open Research Europe},
  volume	= {3},
  pages		= {81},
  year		= {2023},
  doi		= {10.12688/openreseurope.15789.2},
  url		= {https://doi.org/10.12688/openreseurope.15789.2}
}

@Article{	  pcovr,
  title		= {Principal covariates regression: Part I. Theory},
  journal	= {Chemometrics and Intelligent Laboratory Systems},
  volume	= {14},
  number	= {1},
  pages		= {155-164},
  year		= {1992},
  note		= {Proceedings of the 2nd Scandinavian Symposium on
		  Chemometrics},
  issn		= {0169-7439},
  doi		= {https://doi.org/10.1016/0169-7439(92)80100-I},
  url		= {https://www.sciencedirect.com/science/article/pii/016974399280100I},
  author	= {Sijmen {de Jong} and Henk A.L. Kiers}
}

@Misc{		  omat24,
  title		= {Open Materials 2024 (OMat24) Inorganic Materials Dataset
		  and Models},
  author	= {Luis Barroso-Luque and Muhammed Shuaibi and Xiang Fu and
		  Brandon M. Wood and Misko Dzamba and Meng Gao and Ammar
		  Rizvi and C. Lawrence Zitnick and Zachary W. Ulissi},
  year		= {2024},
  eprint	= {2410.12771},
  archiveprefix	= {arXiv},
  primaryclass	= {cond-mat.mtrl-sci},
  url		= {https://arxiv.org/abs/2410.12771}
}

@Misc{		  omol,
  title		= {The Open Molecules 2025 (OMol25) Dataset, Evaluations, and
		  Models},
  author	= {Daniel S. Levine and Muhammed Shuaibi and Evan Walter
		  Clark Spotte-Smith and Michael G. Taylor and Muhammad R.
		  Hasyim and Kyle Michel and Ilyes Batatia and Gábor Csányi
		  and Misko Dzamba and Peter Eastman and Nathan C. Frey and
		  Xiang Fu and Vahe Gharakhanyan and Aditi S. Krishnapriyan
		  and Joshua A. Rackers and Sanjeev Raja and Ammar Rizvi and
		  Andrew S. Rosen and Zachary Ulissi and Santiago Vargas and
		  C. Lawrence Zitnick and Samuel M. Blau and Brandon M.
		  Wood},
  year		= {2025},
  eprint	= {2505.08762},
  archiveprefix	= {arXiv},
  primaryclass	= {physics.chem-ph},
  url		= {https://arxiv.org/abs/2505.08762}
}

@Misc{		  omc25,
  title		= {Open Molecular Crystals 2025 (OMC25) Dataset and Models},
  author	= {Vahe Gharakhanyan and Luis Barroso-Luque and Yi Yang and
		  Muhammed Shuaibi and Kyle Michel and Daniel S. Levine and
		  Misko Dzamba and Xiang Fu and Meng Gao and Xingyu Liu and
		  Haoran Ni and Keian Noori and Brandon M. Wood and Matt
		  Uyttendaele and Arman Boromand and C. Lawrence Zitnick and
		  Noa Marom and Zachary W. Ulissi and Anuroop Sriram},
  year		= {2025},
  eprint	= {2508.02651},
  archiveprefix	= {arXiv},
  primaryclass	= {physics.chem-ph},
  url		= {https://arxiv.org/abs/2508.02651}
}

@Article{	  oc20,
  title		= {Open Catalyst 2020 (OC20) Dataset and Community
		  Challenges},
  volume	= {11},
  issn		= {2155-5435},
  url		= {http://dx.doi.org/10.1021/acscatal.0c04525},
  doi		= {10.1021/acscatal.0c04525},
  number	= {10},
  journal	= {ACS Catalysis},
  publisher	= {American Chemical Society (ACS)},
  author	= {Chanussot, Lowik and Das, Abhishek and Goyal, Siddharth
		  and Lavril, Thibaut and Shuaibi, Muhammed and Riviere,
		  Morgane and Tran, Kevin and Heras-Domingo, Javier and Ho,
		  Caleb and Hu, Weihua and Palizhati, Aini and Sriram,
		  Anuroop and Wood, Brandon and Yoon, Junwoong and Parikh,
		  Devi and Zitnick, C. Lawrence and Ulissi, Zachary},
  year		= {2021},
  month		= may,
  pages		= {6059–6072}
}

@Misc{		  odac,
  title		= {The Open DAC 2023 Dataset and Challenges for Sorbent
		  Discovery in Direct Air Capture},
  author	= {Anuroop Sriram and Sihoon Choi and Xiaohan Yu and Logan M.
		  Brabson and Abhishek Das and Zachary Ulissi and Matt
		  Uyttendaele and Andrew J. Medford and David S. Sholl},
  year		= {2023},
  eprint	= {2311.00341},
  archiveprefix	= {arXiv},
  primaryclass	= {cond-mat.mtrl-sci},
  url		= {https://arxiv.org/abs/2311.00341}
}

@Article{	  mace,
  author	= {Batatia, Ilyes and Benner, Philipp and Chiang, Yuan and
		  Elena, Alin M. and Kov{\'a}cs, D{\'a}vid P. and Riebesell,
		  Janosh and Advincula, Xavier R. and Asta, Mark and Avaylon,
		  Matthew and Baldwin, William J. and Berger, Fabian and
		  Bernstein, Noam and Bhowmik, Arghya and Bigi, Filippo and
		  Blau, Samuel M. and C{\u{a}}rare, Vlad and Ceriotti,
		  Michele and Chong, Sanggyu and Darby, James P. and De,
		  Sandip and Della Pia, Flaviano and Deringer, Volker L. and
		  Elijo{\v{s}}ius, Rokas and El-Machachi, Zakariya and Fako,
		  Edvin and Falcioni, Fabio and Ferrari, Andrea C. and
		  Gardner, John L. A. and Gawkowski, Miko{\l}aj J. and
		  Genreith-Schriever, Annalena and George, Janine and
		  Goodall, Rhys E. A. and Grandel, Jonas and Grey, Clare P.
		  and Grigorev, Petr and Han, Shuang and Handley, Will and
		  Heenen, Hendrik H. and Hermansson, Kersti and Ho, Cheuk Hin
		  and Hofmann, Stephan and Holm, Christian and Jaafar, Jad
		  and Jakob, Konstantin S. and Jung, Hyunwook and Kapil,
		  Venkat and Kaplan, Aaron D. and Karimitari, Nima and
		  Kermode, James R. and Kourtis, Panagiotis and Kroupa, Namu
		  and Kullgren, Jolla and Kuner, Matthew C. and Kuryla,
		  Domantas and Liepuoniute, Guoda and Lin, Chen and Margraf,
		  Johannes T. and Magd{\u{a}}u, Ioan-Bogdan and Michaelides,
		  Angelos and Moore, J. Harry and Naik, Aakash A. and
		  Niblett, Samuel P. and Norwood, Sam Walton and O'Neill,
		  Niamh and Ortner, Christoph and Persson, Kristin A. and
		  Reuter, Karsten and Rosen, Andrew S. and Rosset, Louise A.
		  M. and Schaaf, Lars L. and Schran, Christoph and Shi,
		  Benjamin X. and Sivonxay, Eric and Stenczel, Tam{\'a}s K.
		  and Sutton, Christopher and Svahn, Viktor and Swinburne,
		  Thomas D. and Tilly, Jules and van der Oord, Cas and
		  Vargas, Santiago and Varga-Umbrich, Eszter and Vegge, Tejs
		  and Vondr{\'a}k, Martin and Wang, Yangshuai and Witt,
		  William C. and Wolf, Thomas and Zills, Fabian and
		  Cs{\'a}nyi, G{\'a}bor},
  title		= {A foundation model for atomistic materials chemistry},
  journal	= {The Journal of Chemical Physics},
  volume	= {163},
  number	= {18},
  pages		= {184110},
  year		= {2025},
  month		= nov,
  doi		= {10.1063/5.0297006}
}

@InProceedings{	  pet,
  author	= {Sergey Pozdnyakov and Michele Ceriotti},
  title		= {Smooth, Exact Rotational Symmetrization for Deep Learning
		  on Point Clouds},
  booktitle	= {Advances in Neural Information Processing Systems (NeurIPS
		  2023)},
  year		= {2023},
  doi		= {10.5555/3666122.3669600},
  url		= {https://arxiv.org/abs/2305.19302}
}

@Misc{		  uma,
  title		= {UMA: A Family of Universal Models for Atoms},
  author	= {Brandon M. Wood and Misko Dzamba and Xiang Fu and Meng Gao
		  and Muhammed Shuaibi and Luis Barroso-Luque and Kareem
		  Abdelmaqsoud and Vahe Gharakhanyan and John R. Kitchin and
		  Daniel S. Levine and Kyle Michel and Anuroop Sriram and
		  Taco Cohen and Abhishek Das and Ammar Rizvi and Sushree
		  Jagriti Sahoo and Zachary W. Ulissi and C. Lawrence
		  Zitnick},
  year		= {2025},
  eprint	= {2506.23971},
  archiveprefix	= {arXiv},
  primaryclass	= {cs.LG},
  url		= {https://arxiv.org/abs/2506.23971}
}

@InProceedings{	  esen,
  title		= {Learning Smooth and Expressive Interatomic Potentials for
		  Physical Property Prediction},
  author	= {Fu, Xiang and Wood, Brandon M and Barroso-Luque, Luis and
		  Levine, Daniel S. and Gao, Meng and Dzamba, Misko and
		  Zitnick, C. Lawrence},
  booktitle	= {Proceedings of the 42nd International Conference on
		  Machine Learning},
  pages		= {17875--17893},
  year		= {2025},
  editor	= {Singh, Aarti and Fazel, Maryam and Hsu, Daniel and
		  Lacoste-Julien, Simon and Berkenkamp, Felix and Maharaj,
		  Tegan and Wagstaff, Kiri and Zhu, Jerry},
  volume	= {267},
  series	= {Proceedings of Machine Learning Research},
  month		= {13--19 Jul},
  publisher	= {PMLR},
  url		= {https://proceedings.mlr.press/v267/fu25h.html}
}

@Misc{		  dpa3,
  title		= {A Graph Neural Network for the Era of Large Atomistic
		  Models},
  author	= {Duo Zhang and Anyang Peng and Chun Cai and Wentao Li and
		  Yuanchang Zhou and Jinzhe Zeng and Mingyu Guo and Chengqian
		  Zhang and Bowen Li and Hong Jiang and Tong Zhu and Weile
		  Jia and Linfeng Zhang and Han Wang},
  year		= {2025},
  eprint	= {2506.01686},
  archiveprefix	= {arXiv},
  primaryclass	= {physics.comp-ph},
  url		= {https://arxiv.org/abs/2506.01686}
}

@Article{	  pet-mad-2025,
  author	= {Mazitov, Arslan and Bigi, Filippo and Kellner, Matthias
		  and Pegolo, Paolo and Tisi, Davide and Fraux, Guillaume and
		  Pozdnyakov, Sergey and Loche, Philip and Ceriotti,
		  Michele},
  date		= {2025/11/27},
  doi		= {10.1038/s41467-025-65662-7},
  id		= {Mazitov2025},
  isbn		= {2041-1723},
  journal	= {Nature Communications},
  number	= {1},
  pages		= {10653},
  title		= {PET-MAD as a lightweight universal interatomic potential
		  for advanced materials modeling},
  url		= {https://doi.org/10.1038/s41467-025-65662-7},
  volume	= {16},
  year		= {2025}
}

@Misc{		  openlamv1,
  title		= {LAMBench: A Benchmark for Large Atomistic Models},
  author	= {Anyang Peng and Chun Cai and Mingyu Guo and Duo Zhang and
		  Chengqian Zhang and Wanrun Jiang and Yinan Wang and Antoine
		  Loew and Chengkun Wu and Weinan E and Linfeng Zhang and Han
		  Wang},
  year		= {2025},
  eprint	= {2504.19578},
  archiveprefix	= {arXiv},
  primaryclass	= {physics.comp-ph},
  url		= {https://arxiv.org/abs/2504.19578}
}

@Article{	  mptraj,
  author	= {Deng, Bowen and Zhong, Peichen and Jun, KyuJung and
		  Riebesell, Janosh and Han, Kevin and Bartel, Christopher J.
		  and Ceder, Gerbrand},
  date		= {2023/09/01},
  doi		= {10.1038/s42256-023-00716-3},
  id		= {Deng2023},
  isbn		= {2522-5839},
  journal	= {Nature Machine Intelligence},
  number	= {9},
  pages		= {1031--1041},
  title		= {CHGNet as a pretrained universal neural network potential
		  for charge-informed atomistic modelling},
  url		= {https://doi.org/10.1038/s42256-023-00716-3},
  volume	= {5},
  year		= {2023}
}

@Article{	  cersonsky_pcovr,
  doi		= {10.1088/2632-2153/abfe7c},
  url		= {https://doi.org/10.1088/2632-2153/abfe7c},
  year		= {2021},
  month		= {jul},
  publisher	= {IOP Publishing},
  volume	= {2},
  number	= {3},
  pages		= {035038},
  author	= {Cersonsky, Rose K and Helfrecht, Benjamin A and Engel,
		  Edgar A and Kliavinek, Sergei and Ceriotti, Michele},
  title		= {Improving sample and feature selection with principal
		  covariates regression},
  journal	= {Machine Learning: Science and Technology}
}

@Misc{		  metatrain,
  title		= {Metatensor and Metatomic: Foundational Libraries for
		  Interoperable Atomistic Machine Learning},
  shorttitle	= {Metatensor and Metatomic},
  author	= {Bigi, Filippo and Abbott, Joseph W. and Loche, Philip and
		  Mazitov, Arslan and Tisi, Davide and Langer, Marcel F. and
		  Goscinski, Alexander and Pegolo, Paolo and Chong, Sanggyu
		  and Goswami, Rohit and Chorna, Sofiia and Kellner, Matthias
		  and Ceriotti, Michele and Fraux, Guillaume},
  year		= {2025},
  month		= aug,
  publisher	= {arXiv},
  doi		= {10.48550/arXiv.2508.15704}
}

@Misc{		  matpes,
  title		= {A Foundational Potential Energy Surface Dataset for
		  Materials},
  author	= {Aaron D. Kaplan and Runze Liu and Ji Qi and Tsz Wai Ko and
		  Bowen Deng and Janosh Riebesell and Gerbrand Ceder and
		  Kristin A. Persson and Shyue Ping Ong},
  year		= {2025},
  eprint	= {2503.04070},
  archiveprefix	= {arXiv},
  primaryclass	= {cond-mat.mtrl-sci},
  url		= {https://arxiv.org/abs/2503.04070}
}

@Article{	  spice,
  author	= {Eastman, Peter and Behara, Pavan Kumar and Dotson, David
		  L. and Galvelis, Raimondas and Herr, John E. and Horton,
		  Josh T. and Mao, Yuezhi and Chodera, John D. and Pritchard,
		  Benjamin P. and Wang, Yuanqing and De Fabritiis, Gianni and
		  Markland, Thomas E.},
  date		= {2023/01/04},
  doi		= {10.1038/s41597-022-01882-6},
  id		= {Eastman2023},
  isbn		= {2052-4463},
  journal	= {Scientific Data},
  number	= {1},
  pages		= {11},
  title		= {SPICE, A Dataset of Drug-like Molecules and Peptides for
		  Training Machine Learning Potentials},
  url		= {https://doi.org/10.1038/s41597-022-01882-6},
  volume	= {10},
  year		= {2023}
}

@Misc{		  pet-mad-dos-2025,
  title		= {A universal machine learning model for the electronic
		  density of states},
  author	= {Wei Bin How and Pol Febrer and Sanggyu Chong and Arslan
		  Mazitov and Filippo Bigi and Matthias Kellner and Sergey
		  Pozdnyakov and Michele Ceriotti},
  year		= {2025},
  eprint	= {2508.17418},
  archiveprefix	= {arXiv},
  primaryclass	= {physics.chem-ph},
  url		= {https://arxiv.org/abs/2508.17418}
}

@Article{	  alexandria,
  title		= {Improving machine-learning models in materials science
		  through large datasets},
  journal	= {Materials Today Physics},
  volume	= {48},
  pages		= {101560},
  year		= {2024},
  issn		= {2542-5293},
  doi		= {https://doi.org/10.1016/j.mtphys.2024.101560},
  url		= {https://www.sciencedirect.com/science/article/pii/S2542529324002360},
  author	= {Jonathan Schmidt and Tiago F.T. Cerqueira and Aldo H.
		  Romero and Antoine Loew and Fabian Jäger and Hai-Chen Wang
		  and Silvana Botti and Miguel A.L. Marques}
}

@Article{	  llpr,
  doi		= {10.1088/2632-2153/ad805f},
  url		= {https://doi.org/10.1088/2632-2153/ad805f},
  year		= {2024},
  month		= {oct},
  publisher	= {IOP Publishing},
  volume	= {5},
  number	= {4},
  pages		= {045018},
  author	= {Bigi, Filippo and Chong, Sanggyu and Ceriotti, Michele and
		  Grasselli, Federico},
  title		= {A prediction rigidity formalism for low-cost uncertainties
		  in trained neural networks},
  journal	= {Machine Learning: Science and Technology}
}

@Article{	  uncertainty_nn,
  author	= "Vazquez-Salazar, Luis Itza and Boittier, Eric D. and
		  Meuwly, Markus",
  title		= "Uncertainty quantification for predictions of atomistic
		  neural networks",
  journal	= "Chem. Sci.",
  year		= "2022",
  volume	= "13",
  issue		= "44",
  pages		= "13068-13084",
  publisher	= "The Royal Society of Chemistry",
  doi		= "10.1039/D2SC04056E",
  url		= "http://dx.doi.org/10.1039/D2SC04056E"
}

@Misc{		  mace_multihead,
  title		= {Cross Learning between Electronic Structure Theories for
		  Unifying Molecular, Surface, and Inorganic Crystal
		  Foundation Force Fields},
  author	= {Ilyes Batatia and Chen Lin and Joseph Hart and Elliott
		  Kasoar and Alin M. Elena and Sam Walton Norwood and Thomas
		  Wolf and Gábor Csányi},
  year		= {2025},
  eprint	= {2510.25380},
  archiveprefix	= {arXiv},
  primaryclass	= {physics.chem-ph},
  url		= {https://arxiv.org/abs/2510.25380}
}

@Article{	  gnome,
  author	= {Merchant, Amil and Batzner, Simon and Schoenholz, Samuel
		  S. and Aykol, Muratahan and Cheon, Gowoon and Cubuk, Ekin
		  Dogus},
  date		= {2023/12/01},
  doi		= {10.1038/s41586-023-06735-9},
  id		= {Merchant2023},
  isbn		= {1476-4687},
  journal	= {Nature},
  number	= {7990},
  pages		= {80--85},
  title		= {Scaling deep learning for materials discovery},
  url		= {https://doi.org/10.1038/s41586-023-06735-9},
  volume	= {624},
  year		= {2023}
}

@Article{	  fast_uncertainty,
  author	= {Albert Zhu and Simon Batzner and Albert Musaelian and
		  Boris Kozinsky},
  title		= {Fast uncertainty estimates in deep learning interatomic
		  potentials},
  journal	= {The Journal of Chemical Physics},
  volume	= {158},
  number	= {16},
  pages		= {164111},
  year		= {2023},
  doi		= {10.1063/5.0136574},
  url		= {https://doi.org/10.1063/5.0136574}
}

@Article{	  stability_classification,
  author	= {Bartel, Christopher J. and Trewartha, Amalie and Wang, Qi
		  and Dunn, Alexander and Jain, Anubhav and Ceder, Gerbrand},
  date		= {2020/07/10},
  doi		= {10.1038/s41524-020-00362-y},
  id		= {Bartel2020},
  isbn		= {2057-3960},
  journal	= {npj Computational Materials},
  number	= {1},
  pages		= {97},
  title		= {A critical examination of compound stability predictions
		  from machine-learned formation energies},
  url		= {https://doi.org/10.1038/s41524-020-00362-y},
  volume	= {6},
  year		= {2020}
}

@Article{	  xie2021bayesianff,
  author	= {Xie, Yu and Vandermause, Jonathan and Sun, Lixin and
		  Cepellotti, Andrea and Kozinsky, Boris},
  date		= {2021/03/19},
  doi		= {10.1038/s41524-021-00510-y},
  id		= {Xie2021},
  isbn		= {2057-3960},
  journal	= {npj Computational Materials},
  number	= {1},
  pages		= {40},
  title		= {Bayesian force fields from active learning for simulation
		  of inter-dimensional transformation of stanene},
  url		= {https://doi.org/10.1038/s41524-021-00510-y},
  volume	= {7},
  year		= {2021}
}

@Article{	  matbench,
  author	= {Riebesell, Janosh and Goodall, Rhys E. A. and Benner,
		  Philipp and Chiang, Yuan and Deng, Bowen and Ceder,
		  Gerbrand and Asta, Mark and Lee, Alpha A. and Jain, Anubhav
		  and Persson, Kristin A.},
  date		= {2025/06/01},
  doi		= {10.1038/s42256-025-01055-1},
  id		= {Riebesell2025},
  isbn		= {2522-5839},
  journal	= {Nature Machine Intelligence},
  number	= {6},
  pages		= {836--847},
  title		= {A framework to evaluate machine learning crystal stability
		  predictions},
  url		= {https://doi.org/10.1038/s42256-025-01055-1},
  volume	= {7},
  year		= {2025}
}

@Misc{		  lambench,
  title		= {LAMBench: A Benchmark for Large Atomistic Models},
  author	= {Anyang Peng and Chun Cai and Mingyu Guo and Duo Zhang and
		  Chengqian Zhang and Wanrun Jiang and Yinan Wang and Antoine
		  Loew and Chengkun Wu and Weinan E and Linfeng Zhang and Han
		  Wang},
  year		= {2025},
  eprint	= {2504.19578},
  archiveprefix	= {arXiv},
  primaryclass	= {physics.comp-ph},
  url		= {https://arxiv.org/abs/2504.19578}
}

@Article{	  schmidt2019mlinmaterials,
  author	= {Schmidt, Jonathan and Marques, M{\'a}rio R. G. and Botti,
		  Silvana and Marques, Miguel A. L.},
  date		= {2019/08/08},
  doi		= {10.1038/s41524-019-0221-0},
  id		= {Schmidt2019},
  isbn		= {2057-3960},
  journal	= {npj Computational Materials},
  number	= {1},
  pages		= {83},
  title		= {Recent advances and applications of machine learning in
		  solid-state materials science},
  url		= {https://doi.org/10.1038/s41524-019-0221-0},
  volume	= {5},
  year		= {2019}
}

@Misc{		  mattersim,
  title		= {MatterSim: A Deep Learning Atomistic Model Across
		  Elements, Temperatures and Pressures},
  author	= {Han Yang and Chenxi Hu and Yichi Zhou and Xixian Liu and
		  Yu Shi and Jielan Li and Guanzhi Li and Zekun Chen and
		  Shuizhou Chen and Claudio Zeni and Matthew Horton and
		  Robert Pinsler and Andrew Fowler and Daniel Zügner and
		  Tian Xie and Jake Smith and Lixin Sun and Qian Wang and
		  Lingyu Kong and Chang Liu and Hongxia Hao and Ziheng Lu},
  year		= {2024},
  eprint	= {2405.04967},
  archiveprefix	= {arXiv},
  primaryclass	= {cond-mat.mtrl-sci},
  url		= {https://arxiv.org/abs/2405.04967}
}

@Misc{		  chiang2025mliparena,
  title		= {MLIP Arena: Advancing Fairness and Transparency in Machine
		  Learning Interatomic Potentials via an Open, Accessible
		  Benchmark Platform},
  author	= {Yuan Chiang and Tobias Kreiman and Christine Zhang and
		  Matthew C. Kuner and Elizabeth Weaver and Ishan Amin and
		  Hyunsoo Park and Yunsung Lim and Jihan Kim and Daryl Chrzan
		  and Aron Walsh and Samuel M. Blau and Mark Asta and Aditi
		  S. Krishnapriyan},
  year		= {2025},
  eprint	= {2509.20630},
  archiveprefix	= {arXiv},
  primaryclass	= {physics.chem-ph},
  url		= {https://arxiv.org/abs/2509.20630}
}

@Article{	  kulichenko2024data,
  title		= {Data Generation for Machine Learning Interatomic
		  Potentials and Beyond},
  author	= {Kulichenko, M. and Nebgen, B. and Lubbers, N. and Smith,
		  J. S. and Barros, K. and Allen, A. E. A. and Habib, A. and
		  Shinkle, E. and Fedik, N. and Li, Y. W. and Messerly, R. A.
		  and Tretiak, S.},
  journal	= {Chemical Reviews},
  year		= {2024},
  volume	= {124},
  number	= {24},
  pages		= {13681--13714},
  doi		= {10.1021/acs.chemrev.4c00572},
  url		= {https://doi.org/10.1021/acs.chemrev.4c00572}
}

@Article{	  kulichenko2024uncertainty,
  author	= {Kulichenko, Maksim and Barros, Kipton and Lubbers,
		  Nicholas and Li, Ying Wai and Messerly, Richard and
		  Tretiak, Sergei and Smith, Justin S. and Nebgen, Benjamin},
  date		= {2023/03/01},
  doi		= {10.1038/s43588-023-00406-5},
  id		= {Kulichenko2023},
  isbn		= {2662-8457},
  journal	= {Nature Computational Science},
  number	= {3},
  pages		= {230--239},
  title		= {Uncertainty-driven dynamics for active learning of
		  interatomic potentials},
  url		= {https://doi.org/10.1038/s43588-023-00406-5},
  volume	= {3},
  year		= {2023}
}

@Article{	  jacobs2025,
  title		= {A practical guide to machine learning interatomic
		  potentials – Status and future},
  journal	= {Current Opinion in Solid State and Materials Science},
  volume	= {35},
  pages		= {101214},
  year		= {2025},
  issn		= {1359-0286},
  doi		= {https://doi.org/10.1016/j.cossms.2025.101214},
  url		= {https://www.sciencedirect.com/science/article/pii/S1359028625000014},
  author	= {Ryan Jacobs and Dane Morgan and Siamak Attarian and Jun
		  Meng and Chen Shen and Zhenghao Wu and Clare Yijia Xie and
		  Julia H. Yang and Nongnuch Artrith and Ben Blaiszik and
		  Gerbrand Ceder and Kamal Choudhary and Gabor Csanyi and
		  Ekin Dogus Cubuk and Bowen Deng and Ralf Drautz and Xiang
		  Fu and Jonathan Godwin and Vasant Honavar and Olexandr
		  Isayev and Anders Johansson and Boris Kozinsky and Stefano
		  Martiniani and Shyue Ping Ong and Igor Poltavsky and KJ
		  Schmidt and So Takamoto and Aidan P. Thompson and Julia
		  Westermayr and Brandon M. Wood}
}

@Article{	  vandermause2020,
  author	= {Vandermause, Jonathan and Torrisi, Steven B. and Batzner,
		  Simon and Xie, Yu and Sun, Lixin and Kolpak, Alexie M. and
		  Kozinsky, Boris},
  date		= {2020/03/18},
  doi		= {10.1038/s41524-020-0283-z},
  id		= {Vandermause2020},
  isbn		= {2057-3960},
  journal	= {npj Computational Materials},
  number	= {1},
  pages		= {20},
  title		= {On-the-fly active learning of interpretable Bayesian force
		  fields for atomistic rare events},
  url		= {https://doi.org/10.1038/s41524-020-0283-z},
  volume	= {6},
  year		= {2020}
}

@InProceedings{	  zou2025data,
  title		= {Data Curation for Machine Learning Interatomic Potentials
		  by Determinantal Point Processes},
  author	= {Joanna Zou and Youssef Marzouk},
  booktitle	= {Proceedings of the International Conference on Learning
		  Representations (ICLR)},
  year		= {2025},
  url		= {https://openreview.net/forum?id=PKGP7tg65A}
}

@Article{	  vazquez-salazar2025,
  author	= {Vazquez-Salazar, Luis Itza and K{\"a}ser, Silvan and
		  Meuwly, Markus},
  date		= {2025/02/15},
  doi		= {10.1038/s41524-024-01473-6},
  id		= {Vazquez-Salazar2025},
  isbn		= {2057-3960},
  journal	= {npj Computational Materials},
  number	= {1},
  pages		= {33},
  title		= {Outlier-detection for reactive machine learned potential
		  energy surfaces},
  url		= {https://doi.org/10.1038/s41524-024-01473-6},
  volume	= {11},
  year		= {2025}
}

@InProceedings{	  qu2024,
  author	= {Qu, Eric and Krishnapriyan, Aditi S.},
  booktitle	= {Advances in Neural Information Processing Systems},
  doi		= {10.52202/079017-4412},
  editor	= {A. Globerson and L. Mackey and D. Belgrave and A. Fan and
		  U. Paquet and J. Tomczak and C. Zhang},
  pages		= {139030--139053},
  publisher	= {Curran Associates, Inc.},
  title		= {The Importance of Being Scalable: Improving the Speed and
		  Accuracy of Neural Network Interatomic Potentials Across
		  Chemical Domains},
  url		= {https://proceedings.neurips.cc/paper_files/paper/2024/file/fad8e1915f66161581bb127ccf01092e-Paper-Conference.pdf},
  volume	= {37},
  year		= {2024}
}

@Article{	  schwalbe_koda_2025,
  title		= {Model-free estimation of completeness, uncertainties, and
		  outliers in atomistic machine learning using information
		  theory},
  volume	= {16},
  issn		= {2041-1723},
  url		= {http://dx.doi.org/10.1038/s41467-025-59232-0},
  doi		= {10.1038/s41467-025-59232-0},
  number	= {1},
  journal	= {Nature Communications},
  publisher	= {Springer Science and Business Media LLC},
  author	= {Schwalbe-Koda, Daniel and Hamel, Sebastien and Sadigh,
		  Babak and Zhou, Fei and Lordi, Vincenzo},
  year		= {2025},
  month		= apr
}

@Article{	  guo2022,
  author	= {Guo, Haoyue and Wang, Qian and Urban, Alexander and
		  Artrith, Nongnuch},
  date		= {2022/08/09},
  doi		= {10.1021/acs.chemmater.2c00267},
  isbn		= {0897-4756},
  journal	= {Chemistry of Materials},
  journal1	= {Chemistry of Materials},
  journal2	= {Chem. Mater.},
  month		= {08},
  number	= {15},
  pages		= {6702--6712},
  publisher	= {American Chemical Society},
  title		= {Artificial Intelligence-Aided Mapping of the
		  Structure--Composition--Conductivity Relationships of
		  Glass--Ceramic Lithium Thiophosphate Electrolytes},
  type		= {doi: 10.1021/acs.chemmater.2c00267},
  url		= {https://doi.org/10.1021/acs.chemmater.2c00267},
  volume	= {34},
  year		= {2022},
  year1		= {2022}
}

@Article{	  cheng2020,
  author	= {Cheng, Bingqing and Griffiths, Ryan-Rhys and Wengert,
		  Simon and Kunkel, Christian and Stenczel, Tamas and Zhu,
		  Bonan and Deringer, Volker L. and Bernstein, Noam and
		  Margraf, Johannes T. and Reuter, Karsten and Csanyi,
		  Gabor},
  date		= {2020/09/15},
  doi		= {10.1021/acs.accounts.0c00403},
  isbn		= {0001-4842},
  journal	= {Accounts of Chemical Research},
  journal1	= {Accounts of Chemical Research},
  journal2	= {Acc. Chem. Res.},
  month		= {09},
  number	= {9},
  pages		= {1981--1991},
  publisher	= {American Chemical Society},
  title		= {Mapping Materials and Molecules},
  type		= {doi: 10.1021/acs.accounts.0c00403},
  url		= {https://doi.org/10.1021/acs.accounts.0c00403},
  volume	= {53},
  year		= {2020},
  year1		= {2020}
}

@Misc{		  deng2024,
  title		= {Overcoming systematic softening in universal machine
		  learning interatomic potentials by fine-tuning},
  author	= {Bowen Deng and Yunyeong Choi and Peichen Zhong and Janosh
		  Riebesell and Shashwat Anand and Zhuohan Li and KyuJung Jun
		  and Kristin A. Persson and Gerbrand Ceder},
  year		= {2024},
  eprint	= {2405.07105},
  archiveprefix	= {arXiv},
  primaryclass	= {cond-mat.mtrl-sci},
  url		= {https://arxiv.org/abs/2405.07105}
}

@Article{	  radova2025,
  author	= {Radova, Mariia and Stark, Wojciech G. and Allen, Connor S.
		  and Maurer, Reinhard J. and Bart{\'o}k, Albert P.},
  date		= {2025/07/18},
  doi		= {10.1038/s41524-025-01727-x},
  id		= {Radova2025},
  isbn		= {2057-3960},
  journal	= {npj Computational Materials},
  number	= {1},
  pages		= {237},
  title		= {Fine-tuning foundation models of materials interatomic
		  potentials with frozen transfer learning},
  url		= {https://doi.org/10.1038/s41524-025-01727-x},
  volume	= {11},
  year		= {2025}
}

@Article{	  kaur2025,
  author	= "Kaur, Harveen and Della Pia, Flaviano and Batatia, Ilyes
		  and Advincula, Xavier R. and Shi, Benjamin X. and Lan,
		  Jinggang and Csányi, Gábor and Michaelides, Angelos and
		  Kapil, Venkat",
  title		= "Data-efficient fine-tuning of foundational models for
		  first-principles quality sublimation enthalpies",
  journal	= "Faraday Discuss.",
  year		= "2025",
  volume	= "256",
  issue		= "0",
  pages		= "120-138",
  publisher	= "The Royal Society of Chemistry",
  doi		= "10.1039/D4FD00107A",
  url		= "http://dx.doi.org/10.1039/D4FD00107A"
}

@Article{	  janet2019chemsci,
  author	= "Janet, Jon Paul and Duan, Chenru and Yang, Tzuhsiung and
		  Nandy, Aditya and Kulik, Heather J.",
  title		= "A quantitative uncertainty metric controls error in neural
		  network-driven chemical discovery",
  journal	= "Chem. Sci.",
  year		= "2019",
  volume	= "10",
  issue		= "34",
  pages		= "7913-7922",
  publisher	= "The Royal Society of Chemistry",
  doi		= "10.1039/C9SC02298H",
  url		= "http://dx.doi.org/10.1039/C9SC02298H"
}

@Article{	  shnorb,
  author	= {Sch{\"u}tt, K. T. and Gastegger, M. and Tkatchenko, A. and
		  M{\"u}ller, K. -R. and Maurer, R. J.},
  date		= {2019/11/15},
  doi		= {10.1038/s41467-019-12875-2},
  id		= {Sch{\"u}tt2019},
  isbn		= {2041-1723},
  journal	= {Nature Communications},
  number	= {1},
  pages		= {5024},
  title		= {Unifying machine learning and quantum chemistry with a
		  deep neural network for molecular wavefunctions},
  url		= {https://doi.org/10.1038/s41467-019-12875-2},
  volume	= {10},
  year		= {2019}
}

@Article{	  suman2025,
  author	= {Suman, Divya and Nigam, Jigyasa and Saade, Sandra and
		  Pegolo, Paolo and T{\"u}rk, Hanna and Zhang, Xing and Chan,
		  Garnet Kin-Lic and Ceriotti, Michele},
  date		= {2025/07/08},
  doi		= {10.1021/acs.jctc.5c00522},
  isbn		= {1549-9618},
  journal	= {Journal of Chemical Theory and Computation},
  journal1	= {Journal of Chemical Theory and Computation},
  journal2	= {J. Chem. Theory Comput.},
  month		= {07},
  number	= {13},
  pages		= {6505--6516},
  publisher	= {American Chemical Society},
  title		= {Exploring the Design Space of Machine Learning Models for
		  Quantum Chemistry with a Fully Differentiable Framework},
  type		= {doi: 10.1021/acs.jctc.5c00522},
  url		= {https://doi.org/10.1021/acs.jctc.5c00522},
  volume	= {21},
  year		= {2025},
  year1		= {2025}
}

@Article{	  how2025,
  title		= {Adaptive energy reference for machine-learning models of
		  the electronic density of states},
  author	= {How, Wei Bin and Chong, Sanggyu and Grasselli, Federico
		  and Huguenin-Dumittan, Kevin K. and Ceriotti, Michele},
  journal	= {Phys. Rev. Mater.},
  volume	= {9},
  issue		= {1},
  pages		= {013802},
  numpages	= {10},
  year		= {2025},
  month		= {Jan},
  publisher	= {American Physical Society},
  doi		= {10.1103/PhysRevMaterials.9.013802},
  url		= {https://link.aps.org/doi/10.1103/PhysRevMaterials.9.013802}
}

@Article{	  mahmoud2020,
  title		= {Learning the electronic density of states in condensed
		  matter},
  author	= {Ben Mahmoud, Chiheb and Anelli, Andrea and Cs\'anyi,
		  G\'abor and Ceriotti, Michele},
  journal	= {Phys. Rev. B},
  volume	= {102},
  issue		= {23},
  pages		= {235130},
  numpages	= {14},
  year		= {2020},
  month		= {Dec},
  publisher	= {American Physical Society},
  doi		= {10.1103/PhysRevB.102.235130},
  url		= {https://link.aps.org/doi/10.1103/PhysRevB.102.235130}
}

@Article{	  feldmann2025,
  author	= "Feldmann, Christian W. and Sieg, Jochen and Mathea,
		  Miriam",
  title		= "Analysis of uncertainty of neural fingerprint-based
		  models",
  journal	= "Faraday Discuss.",
  year		= "2025",
  volume	= "256",
  issue		= "0",
  pages		= "551-567",
  publisher	= "The Royal Society of Chemistry",
  doi		= "10.1039/D4FD00095A",
  url		= "http://dx.doi.org/10.1039/D4FD00095A"
}

@Misc{		  grace,
  title		= {Graph atomic cluster expansion for foundational machine
		  learning interatomic potentials},
  author	= {Yury Lysogorskiy and Anton Bochkarev and Ralf Drautz},
  year		= {2025},
  eprint	= {2508.17936},
  archiveprefix	= {arXiv},
  primaryclass	= {cond-mat.mtrl-sci},
  url		= {https://arxiv.org/abs/2508.17936}
}

@Misc{		  orb3,
  title		= {Orb-v3: atomistic simulation at scale},
  author	= {Benjamin Rhodes and Sander Vandenhaute and Vaidotas
		  Šimkus and James Gin and Jonathan Godwin and Tim Duignan
		  and Mark Neumann},
  year		= {2025},
  eprint	= {2504.06231},
  archiveprefix	= {arXiv},
  primaryclass	= {cond-mat.mtrl-sci},
  url		= {https://arxiv.org/abs/2504.06231}
}

@Article{	  jacobs1991adaptive,
  title		= {Adaptive Mixtures of Local Experts},
  author	= {Jacobs, Robert A. and Jordan, Michael I. and Nowlan,
		  Steven J. and Hinton, Geoffrey E.},
  journal	= {Neural Computation},
  volume	= {3},
  number	= {1},
  pages		= {79--87},
  year		= {1991},
  doi		= {10.1162/neco.1991.3.1.79},
  url		= {https://doi.org/10.1162/neco.1991.3.1.79}
}

@InProceedings{	  riquelme2021scaling,
  author	= {Riquelme, Carlos and Puigcerver, Joan and Mustafa, Basil
		  and Neumann, Maxim and Jenatton, Rodolphe and Susano Pinto,
		  Andr{\'e} and Keysers, Daniel and Houlsby, Neil},
  title		= {Scaling Vision with Sparse Mixture of Experts},
  booktitle	= {Advances in Neural Information Processing Systems},
  volume	= {34},
  pages		= {8583--8595},
  year		= {2021}
}

@InProceedings{	  shazeer2017outrageously,
  title		= {Outrageously Large Neural Networks: The Sparsely-Gated
		  Mixture-of-Experts Layer},
  author	= {Shazeer, Noam and Mirhoseini, Azalia and Maziarz,
		  Krzysztof and Davis, Andy and Le, Quoc V. and Hinton,
		  Geoffrey and Dean, Jeff},
  booktitle	= {Proceedings of the International Conference on Learning
		  Representations (ICLR)},
  year		= {2017},
  url		= {https://openreview.net/forum?id=B1ckMDqlg}
}

@Article{	  caruana1997,
  author	= {Caruana, Rich},
  date		= {1997/07/01},
  doi		= {10.1023/A:1007379606734},
  id		= {Caruana1997},
  isbn		= {1573-0565},
  journal	= {Machine Learning},
  number	= {1},
  pages		= {41--75},
  title		= {Multitask Learning},
  url		= {https://doi.org/10.1023/A:1007379606734},
  volume	= {28},
  year		= {1997}
}

@Misc{		  ruder2017,
  title		= {An Overview of Multi-Task Learning in Deep Neural
		  Networks},
  author	= {Sebastian Ruder},
  year		= {2017},
  eprint	= {1706.05098},
  archiveprefix	= {arXiv},
  primaryclass	= {cs.LG},
  url		= {https://arxiv.org/abs/1706.05098}
}

@Article{	  deepmd-kit-v3,
  author	= { Jinzhe Zeng and Duo Zhang and Anyang Peng and Xiangyu
		  Zhang and Sensen He and Yan Wang and Xinzijian Liu and
		  Hangrui Bi and Yifan Li and Chun Cai and Chengqian Zhang
		  and Yiming Du and Jia-Xin Zhu and Pinghui Mo and Zhengtao
		  Huang and Qiyu Zeng and Shaochen Shi and Xuejian Qin and
		  Zhaoxi Yu and Chenxing Luo and Ye Ding and Yun-Pei Liu and
		  Ruosong Shi and Zhenyu Wang and Sigbj{\o}rn L{\o}land Bore
		  and Junhan Chang and Zhe Deng and Zhaohan Ding and Siyuan
		  Han and Wanrun Jiang and Guolin Ke and Zhaoqing Liu and
		  Denghui Lu and Koki Muraoka and Hananeh Oliaei and Anurag
		  Kumar Singh and Haohui Que and Weihong Xu and Zhangmancang
		  Xu and Yong-Bin Zhuang and Jiayu Dai and Timothy J. Giese
		  and Weile Jia and Ben Xu and Darrin M. York and Linfeng
		  Zhang and Han Wang },
  title		= { {DeePMD-kit v3: A Multiple-Backend Framework for Machine
		  Learning Potentials} },
  journal	= {J. Chem. Theory Comput.},
  year		= 2025,
  volume	= 21,
  number	= 9,
  pages		= {4375--4385},
  doi		= {10.1021/acs.jctc.5c00340}
}

@Article{	  maceoff,
  author	= {Kov{\'a}cs, D{\'a}vid P{\'e}ter and Moore, J. Harry and
		  Browning, Nicholas J. and Batatia, Ilyes and Horton, Joshua
		  T. and Pu, Yixuan and Kapil, Venkat and Witt, William C.
		  and Magd{\u a}u, Ioan-Bogdan and Cole, Daniel J. and
		  Cs{\'a}nyi, G{\'a}bor},
  date		= {2025/05/28},
  doi		= {10.1021/jacs.4c07099},
  isbn		= {0002-7863},
  journal	= {Journal of the American Chemical Society},
  journal1	= {Journal of the American Chemical Society},
  journal2	= {J. Am. Chem. Soc.},
  month		= {05},
  number	= {21},
  pages		= {17598--17611},
  publisher	= {American Chemical Society},
  title		= {MACE-OFF: Short-Range Transferable Machine Learning Force
		  Fields for Organic Molecules},
  type		= {doi: 10.1021/jacs.4c07099},
  url		= {https://doi.org/10.1021/jacs.4c07099},
  volume	= {147},
  year		= {2025},
  year1		= {2025}
}

@Article{	  furness2020,
  author	= {Furness, James W. and Kaplan, Aaron D. and Ning, Jinliang
		  and Perdew, John P. and Sun, Jianwei},
  date		= {2020/10/01},
  doi		= {10.1021/acs.jpclett.0c02405},
  journal	= {The Journal of Physical Chemistry Letters},
  journal1	= {The Journal of Physical Chemistry Letters},
  journal2	= {J. Phys. Chem. Lett.},
  month		= {10},
  number	= {19},
  pages		= {8208--8215},
  publisher	= {American Chemical Society},
  title		= {Accurate and Numerically Efficient r2SCAN Meta-Generalized
		  Gradient Approximation},
  type		= {doi: 10.1021/acs.jpclett.0c02405},
  url		= {https://doi.org/10.1021/acs.jpclett.0c02405},
  volume	= {11},
  year		= {2020},
  year1		= {2020}
}

@Article{	  glielmo2022,
  author	= {Glielmo, Aldo and Zeni, Claudio and Cheng, Bingqing and
		  Cs{\'a}nyi, G{\'a}bor and Laio, Alessandro},
  title		= {Ranking the information content of distance measures},
  journal	= {PNAS Nexus},
  volume	= {1},
  number	= {2},
  pages		= {pgac039},
  year		= {2022},
  doi		= {10.1093/pnasnexus/pgac039},
  issn		= {2752-6542}
}

@Misc{		  matcloud_record,
  title		= {Comparing the latent features of universal
		  machine-learning interatomic potentials},
  author	= {Sofiia Chorna and Davide Tisi and Cesare Malosso and Wei
		  Bin How and Michele Ceriotti},
  year		= {2025},
  publisher	= {Materials Cloud},
  howpublished	= {https://doi.org/10.24435/materialscloud:gb-5z},
  doi		= {10.24435/materialscloud:gb-5z}
}

@Article{	  talirz_2020,
  title		= {Materials Cloud, a platform for open computational
		  science},
  volume	= {7},
  issn		= {2052-4463},
  url		= {http://dx.doi.org/10.1038/s41597-020-00637-5},
  doi		= {10.1038/s41597-020-00637-5},
  number	= {1},
  journal	= {Scientific Data},
  publisher	= {Springer Science and Business Media LLC},
  author	= {Talirz, Leopold and Kumbhar, Snehal and Passaro, Elsa and
		  Yakutovich, Aliaksandr V. and Granata, Valeria and
		  Gargiulo, Fernando and Borelli, Marco and Uhrin, Martin and
		  Huber, Sebastiaan P. and Zoupanos, Spyros and Adorf, Carl
		  S. and Andersen, Casper Welzel and Schütt, Ole and
		  Pignedoli, Carlo A. and Passerone, Daniele and
		  VandeVondele, Joost and Schulthess, Thomas C. and Smit,
		  Berend and Pizzi, Giovanni and Marzari, Nicola},
  year		= {2020},
  month		= sep
}

@Misc{		  li2025platonicrepresentation,
  title		= {Platonic representation of foundation machine learning
		  interatomic potentials},
  author	= {Zhenzhu Li and Aron Walsh},
  year		= {2025},
  eprint	= {2512.05349},
  archiveprefix	= {arXiv},
  primaryclass	= {cond-mat.mtrl-sci},
  url		= {https://arxiv.org/abs/2512.05349}
}

@Misc{		  bombarelli2025,
  title		= {Universally Converging Representations of Matter Across
		  Scientific Foundation Models},
  author	= {Sathya Edamadaka and Soojung Yang and Ju Li and Rafael
		  Gómez-Bombarelli},
  year		= {2025},
  eprint	= {2512.03750},
  archiveprefix	= {arXiv},
  primaryclass	= {cs.LG},
  url		= {https://arxiv.org/abs/2512.03750}
}

@Misc{		  upet,
  title		= {{UPET}: Universal interatomic potentials for advanced
		  materials modeling},
  author	= {{Laboratory of Computational Science and Modeling (COSMO),
		  EPFL}},
  year		= {2026},
  howpublished	= {\url{https://github.com/lab-cosmo/upet}},
  note		= {Online; accessed Jan.\ 2026}
}

@Article{	  muller2025peeringinside,
  author	= {Bonneau, Klara and Lederer, Jonas and Templeton, Clark and
		  Rosenberger, David and Giambagli, Lorenzo and M{\"u}ller,
		  Klaus-Robert and Clementi, Cecilia},
  date		= {2025/11/10},
  doi		= {10.1038/s41467-025-65863-0},
  id		= {Bonneau2025},
  isbn		= {2041-1723},
  journal	= {Nature Communications},
  number	= {1},
  pages		= {9898},
  title		= {Peering inside the black box by learning the relevance of
		  many-body functions in neural network potentials},
  url		= {https://doi.org/10.1038/s41467-025-65863-0},
  volume	= {16},
  year		= {2025}
}

\clearpage

\onecolumngrid
\part*{Supplementary Information}

\setcounter{figure}{0}
\renewcommand{\thefigure}{S\arabic{figure}}

\setcounter{table}{0}
\renewcommand{\thetable}{S\arabic{table}}

\setcounter{equation}{0}
\renewcommand{\theequation}{S\arabic{equation}}

\section{Feature space comparison metrics}
\label{sec:si-errors}

\begin{figure}[h!]
    \centering
    \includegraphics[width=\linewidth]{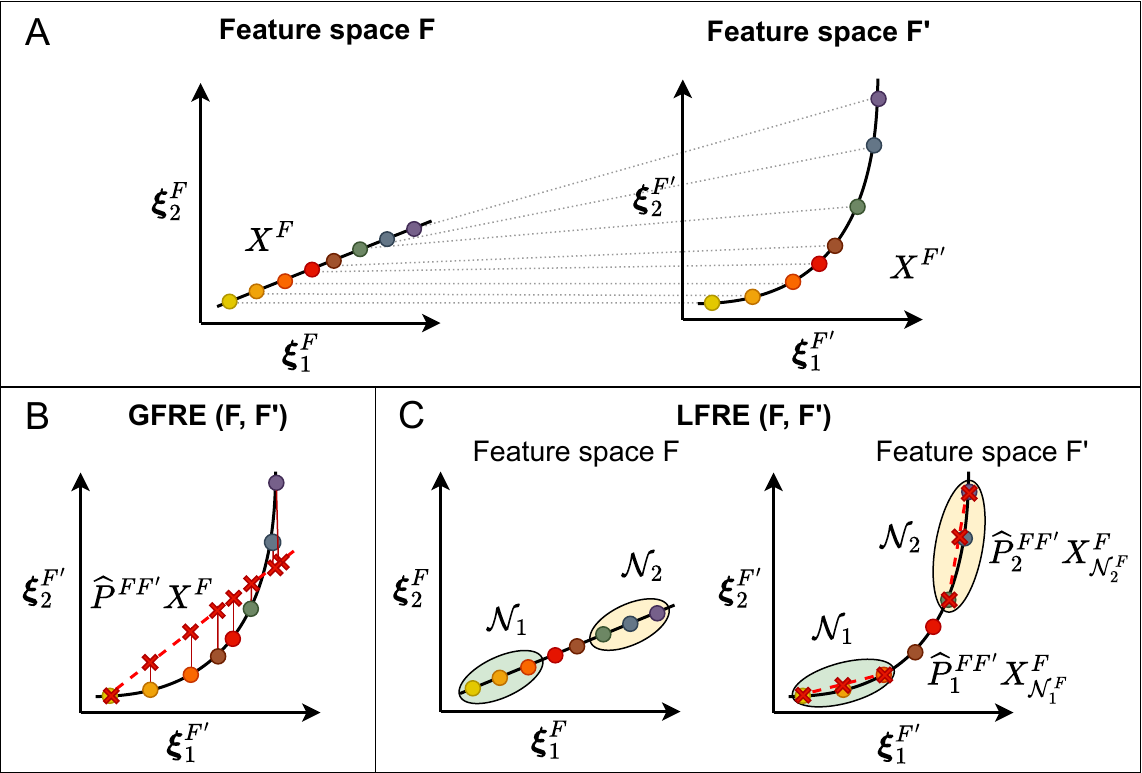}
    \caption{A schematic overview of the global (GFRE) and local (LFRE) feature reconstruction errors. (A) shows a one-dimensional manifold of atomic environments embedded into two
    two-dimensional feature spaces, $F$ and $F'$, related by a nonlinear transformation. (B) illustrates $\mathrm{GFRE}(F, F')$, where a single global linear mapping
    $\widehat{P}^{FF'}$ is applied to reconstruct $F'$ from $F$. The dashed red line indicates the
    globally optimal linear fit $\widehat{P}^{FF'}X^F$. (C) shows $\mathrm{LFRE}(F, F')$: for each atomic environment $i$, a separate local linear mapping $\widehat{P}^{FF'}_{(i)}$ is fitted to its $k$ nearest neighbors $\mathcal{N}_i$ in feature space $F$.}
    \label{fig:errors}
\end{figure}

To aid the reader's understanding of the the difference between the global feature reconstruction error (GFRE) and the local feature reconstruction error (LFRE) introduced in Sec.~\ref{sec:reconstruction-errors}, we present a minimal illustrative example in Fig. \ref{fig:errors}.

\section{Latent feature comparison between uMLIPs on Alexandria dataset}
\label{sec:si-errors-umlip-alexandria}

\begin{figure}[h!]
    \centering
    \includegraphics[width=\linewidth]{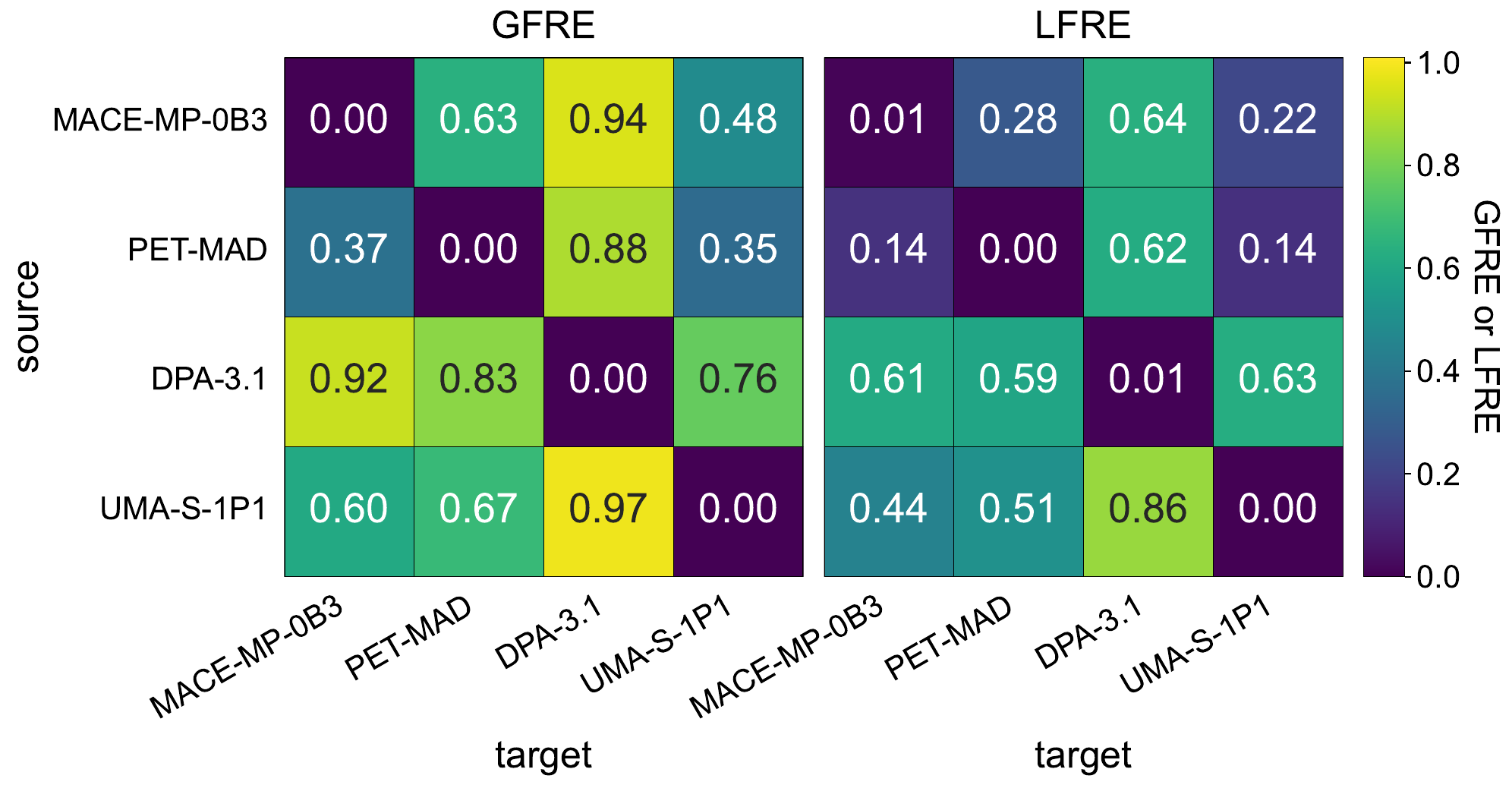}
    \caption{The reconstruction errors for atomic last-layer latent features of MACE-MP-03b, PET-MAD, DPA-3.1, and UMA-S-1P1, computed for the atomic environments sampled from the sAlex dataset.}
    \label{fig:errors_alexandria}
\end{figure}

To ensure independence from training distributions and mitigate potential biases, we evaluate reconstruction errors on the subsampled Alexandria dataset (sAlex) \cite{omat24, Alexandria}, which contains a broad coverage of inorganic materials across three-, two-, and one-dimensional periodic structures. We first randomly select 20,000 structures from the validation split, then filter to exclude the single-atom structures and structures with no interatomic distances below 6~\AA ~(incompatible with UMA-S-1P1), and those containing actinide elements (as atomic numbers 89--94 are not supported by MACE-MP-0b3), yielding 169,836 atomic environments from 16,306 structures. From these, 1000 atomic environments are randomly sampled to comprise $\mathcal{D}_{\mathrm{test}}$, and all remaining environments are used as $\mathcal{D}_{\mathrm{train}}$ to obtain the regression weights $\widehat{P}_{FF'}$.

Similarly to the results on MAD test (see Fig. \ref{fig:umlip-errors} in the main text), PET-MAD reconstructs other models' feature spaces with the lowest GFRE and LFRE (on average 0.53 and 0.3, respectively), as shown in Fig.~\ref{fig:errors_alexandria}. The high incoming and outgoing errors for DPA-3.1 suggest that its latent space is particularly orthogonal to other models.

\section{Comparing backbone features of uMLIPs}
\label{sec:si-errors-umlip-bb}

The lower reconstruction errors for backbone features (average off-diagonal GFRE 0.54 vs 0.66 for last-layer, LFRE 0.30 vs 0.37) indicate that representations immediately after message-passing are more consistent and linearly reconstructable.

\begin{figure}[h!]
    \centering
    \includegraphics[width=\linewidth]{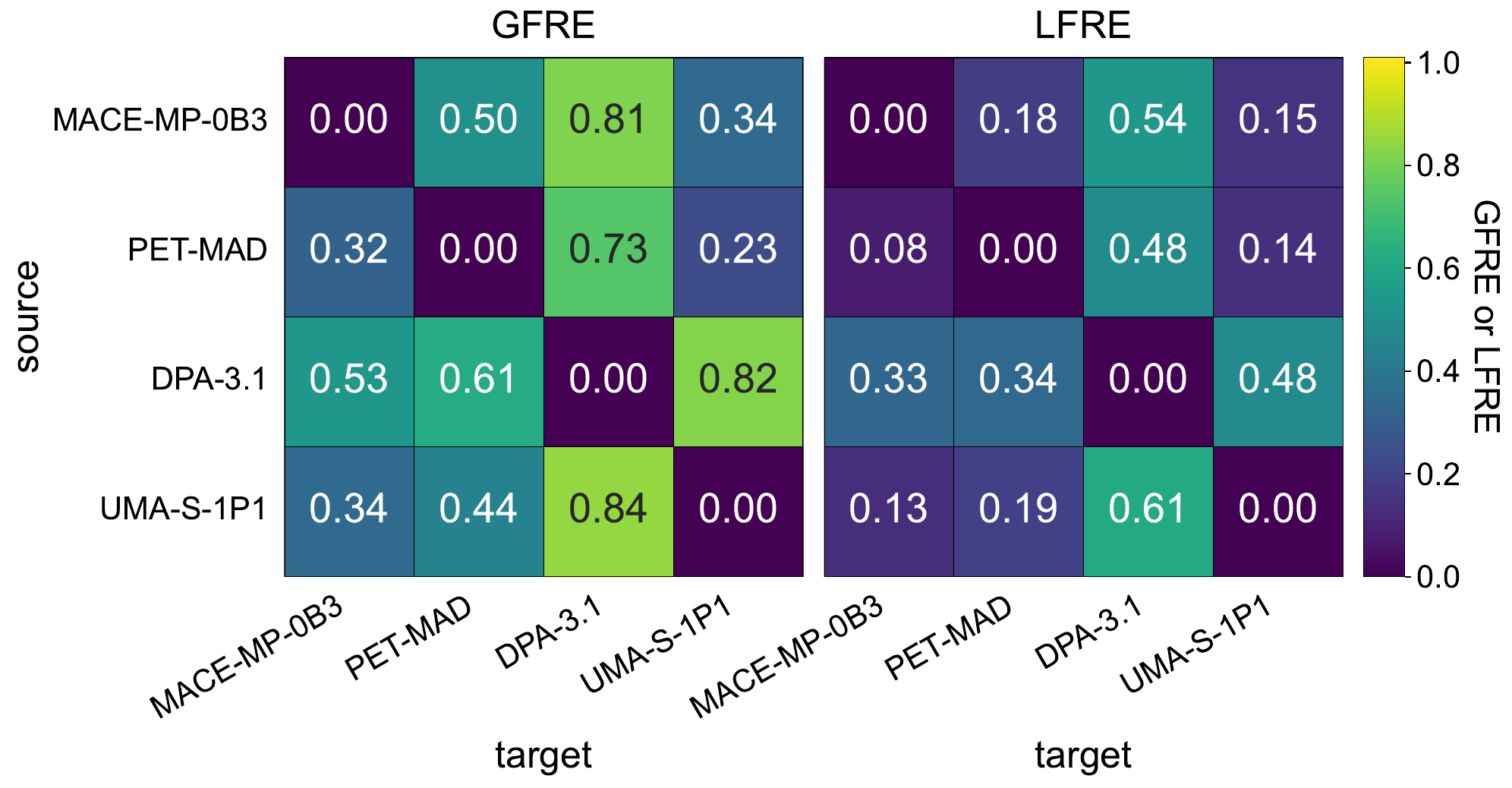}
    \caption{The reconstruction errors for atomic backbone features computed across MACE-MP-0b3, PET-MAD, DPA-3.1, and UMA-S-1P1, evaluated on the MAD test subset. Compared to last-layer features (Fig. \ref{fig:umlip-errors}), backbone representations exhibit consistently lower cross-model errors.
    }
    \label{fig:umlip-errors-bb}
\end{figure}

\section{Comparing model variants}
\label{sec:si-model-variants}

\subsection{Multi-task handling}

\subsubsection{DPA-3.1}
\label{sec:si-variants-dpa}

\begin{figure}[h!]
    \centering
    \includegraphics[width=\linewidth]{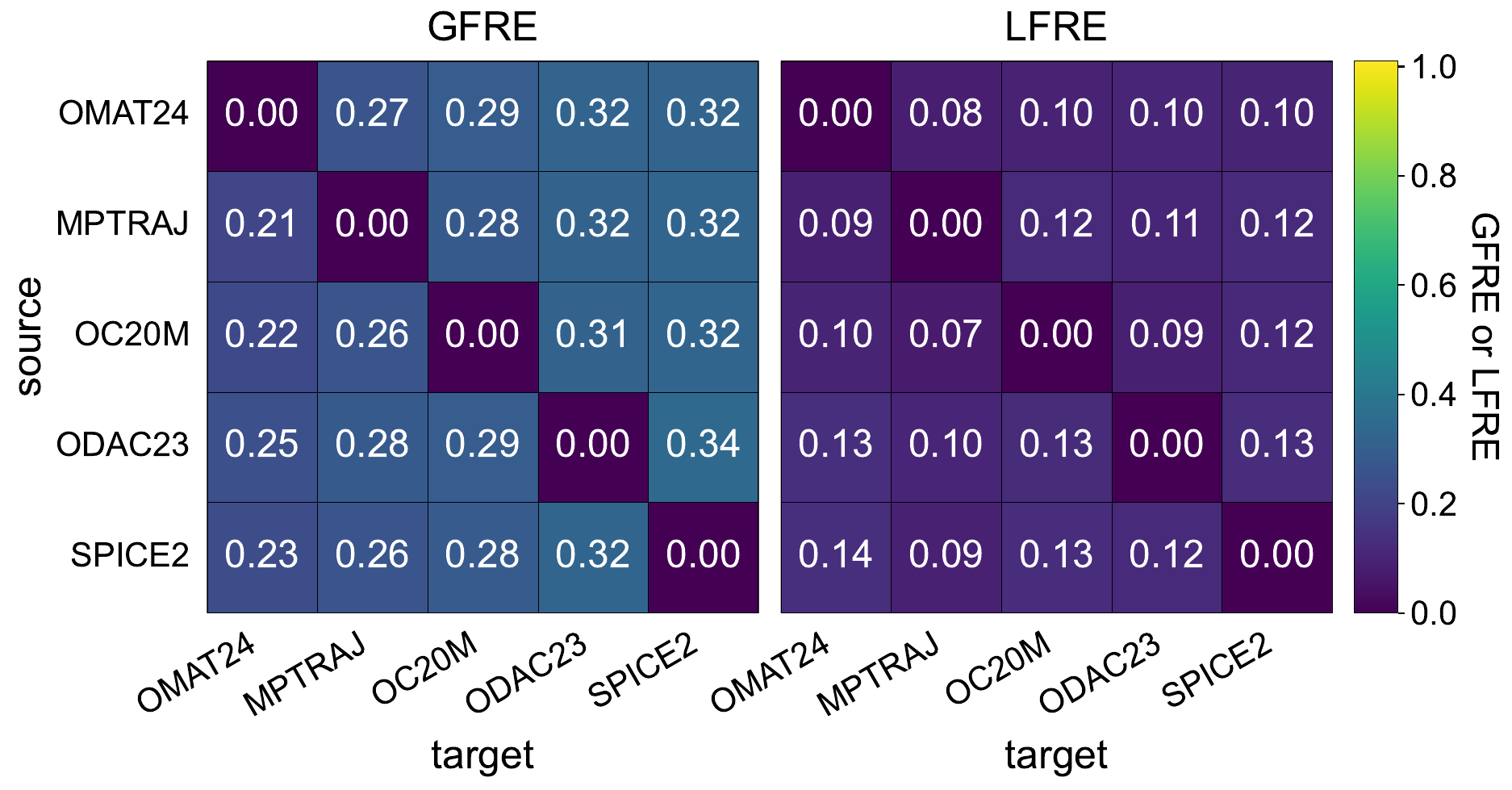}
    \caption{Reconstruction errors among DPA-3.1 branches trained on OMat24, MPtrj, OC20, ODAC, and SPICE.}
    \label{fig:variants-dpa_errors}
\end{figure}

DPA-3.1 incorporates a multi-task framework with a shared backbone network conditioned on dataset identifiers via one-hot encoding \cite{DPA3}. We compare its branches corresponding to OMat24, MPtrj, OC20, ODAC, and SPICE \cite{omat24, MPTRAJ, OC20, ODAC, SPICE}. As shown in Fig.~\ref{fig:variants-dpa_errors}, the latent spaces of these branches are remarkably consistent despite the extreme diversity of the underlying data. LFRE remains particularly low (with the maximum of 0.14), indicating highly similar local neighborhoods across branches. This suggests that this multi-task design with full backbone sharing enforces highly similar latent representations across branches.

\subsubsection{Universal PET checkpoints}
\label{sec:si-variants-pet}

\begin{figure}[h!]
    \centering
    \includegraphics[width=\linewidth]{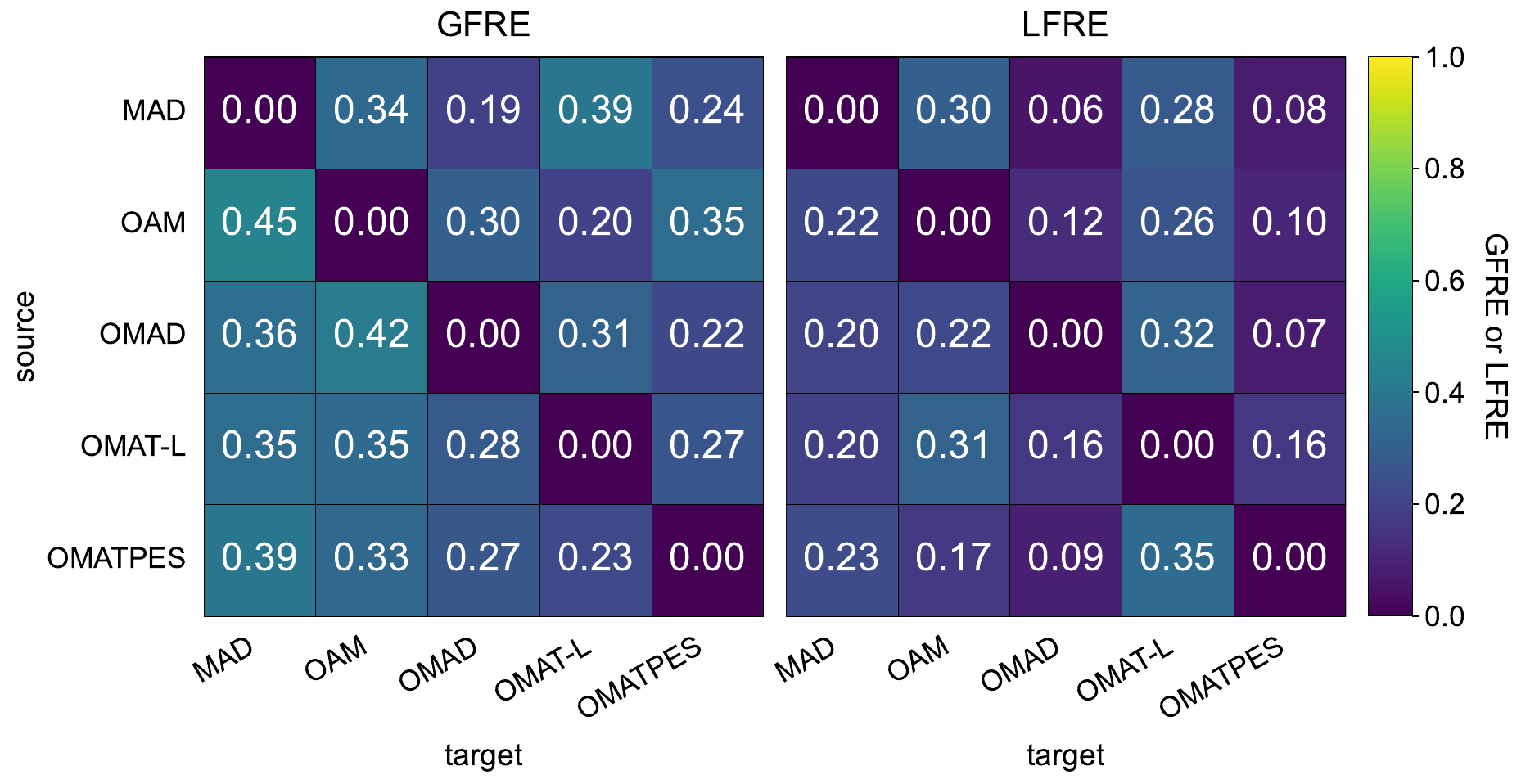}
    \caption{Reconstruction errors among PET variants trained on MAD, OAM, OMAT, and OMATPES on the MAD test \cite{omat24, MATPES, MPTRAJ, MAD}.}
    \label{fig:variants-errors_pet}
\end{figure}

Here, we evaluate the feature reconstruction errors across single-task PET universal models trained on different datasets. Table S1 summarizes the  characteristics of the five PET variants analyzed, including model size and training protocol. All models were obtained from the public UPET repository \cite{uPET}. From Fig.~\ref{fig:variants-errors_pet}, MAD and OAM checkpoints are the most challenging to reconstruct with global linear approximation. OMat24 pretrained models (OMAD, OMAT, and OMATPES) show slightly lower reconstruction, sharing the same latent space foundation from the pre-training.

\begin{table}[h!]
    \centering
    \label{tab:pet-variants}
    \caption{Overview of PET model variants used in the reconstruction error analysis. 
    All ``O''-prefixed models are pre-trained on OMat24 before training on the target dataset. sAlex stands for the subsampled Alexandria dataset (sAlexandria) \cite{omat24, Alexandria}.}    
    \begin{tabular}{c cccc}
        \hline
        Model & Pre-training & Target Dataset & DFT Level & Size \\
        \hline
        PET-MAD & — & MAD & PBEsol & S \\
        PET-OAM & OMat24 & sAlexandria + MPtrj & PBE & L \\
        PET-OMAD & OMat24 & MAD & PBEsol & L \\
        PET-OMAT & OMat24 & OMat24 & PBE & L \\
        PET-OMATPES & OMat24 & MatPES & r2SCAN & L \\
        \hline
    \end{tabular}
\end{table}

\subsection{Dataset domain}
\label{sec:si-variants-domains}

We further probe the effect of training dataset domain by comparing medium-sized models, MACE-MP-0b3 \cite{MACE} trained on MPtrj \cite{MPTRAJ}, a materials-focused dataset, and MACE-OFF23 \cite{MACEOFF} trained on SPICE \cite{SPICE}, an organics-focused dataset.

Reconstruction errors have been computed for only the organic structures found within the MAD test set. From Table \ref{tab:mace-domains}, MACE-MP-0b3 can reconstruct the organic-trained model almost perfectly (GFRE = 0.02). In the reverse direction, the MACE-OFF23 fails to reconstruct the MACE-MP-0b3 features globally (GFRE = 0.36), despite near-perfect local feature reconstruction (LFRE = 0.04). This is explained by the fact that MACE-MP-0b3 was trained on a vastly more diverse dataset, and has learned a richer global mapping that MACE-OFF23 simply cannot replicate due to a significantly smaller dataset scope.

\begin{table}[h!]
    \centering
    \caption{GFRE and LFRE comparison between MACE-MP-03b and MACE-OFF23 evaluated on the organic structures of the MAD test set.}    
    \begin{tabular}{c cc}
        \hline
        Reconstruction direction (source  $\rightarrow$ target)  & GFRE & LFRE \\
        \hline
        MP-03b $\rightarrow$ OFF23 & 0.02 & 0.06 \\
        OFF23 $\rightarrow$ MP-03b & 0.36 & 0.04 \\
        \hline
    \end{tabular}
    \label{tab:mace-domains}
\end{table}

\subsection{Model sizes}
\label{sec:si-variants-sizes}

\subsubsection{MACE-MP-0a and MACE-OFF23}
\label{sec:si-variants-sizes-mace}

\begin{figure}[h!]
    \centering
    \includegraphics[width=\linewidth]{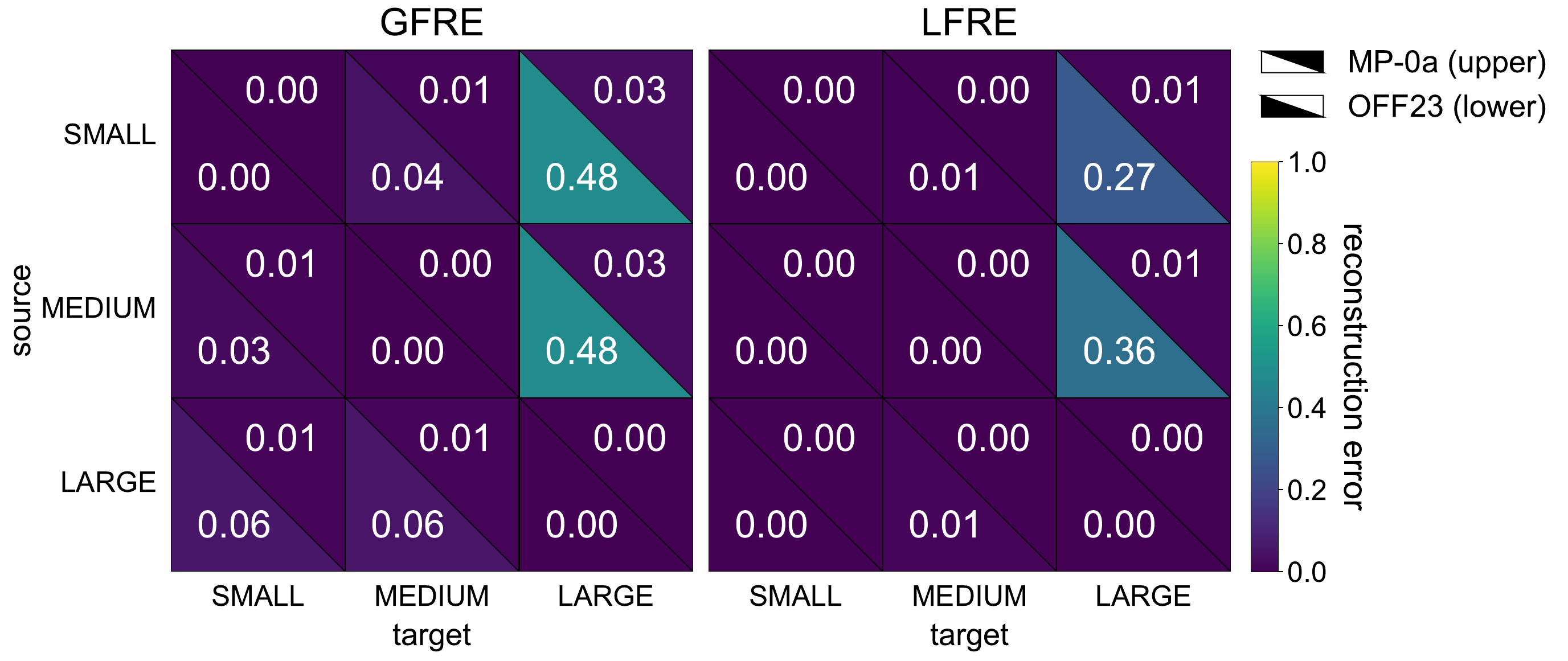}
    \caption{GFRE (left) and LFRE (right) among MACE checkpoints of different sizes (small, medium, large) on the MAD test dataset. Upper triangles correspond to MACE-MP-0a (materials); lower to MACE-OFF23 (organics).}
    \label{fig:mace-sizes}
\end{figure}

In this section, we examine how well the MACE models of different sizes (small, medium, large) can reconstruct the last-layer features of one another. In Fig. \ref{fig:mace-sizes}, the upper triangle shows results for the MPtrj-trained MACE-MP-0a evaluated on the MAD test subset, and the lower triangle shows the results for MACE-OFF23 \cite{MACEOFF} evaluated on the organic structures found in the MAD test subset.

Notably, MACE-MP-0a checkpoints reconstruct each other almost perfectly, regardless of the model size. For MACE-OFF23, error rises when a small or medium model tries to reconstruct the large one (GFRE is 0.48, LFRE is 0.31 on average between them), meaning the larger model contains expressive capacity that smaller models cannot fully replicate. These distinct trends could stem from the differences in the hyperparameters or training protocol between MACE-MP series and MACE-OFF, and/or the differences in the scope of the training datasets.

\subsubsection{PET-OMAT}
\label{sec:si-variants-sizes-pet}

In addition to MACE variants, we also compare the PET-OMAT checkpoints \cite{uPET} of different sizes. The model sizes depend on the combination of the number of GNN layers ($N_{\text{gnn}}$), the number of transformer layers ($N_{\text{tl}}$), the neural network width ($d_{\text{pet}}$), and the cutoff radius ($r_{\text{cut}}$) (see Table~\ref{tab:pet-omat}).

\begin{table}[tbh]
    \centering
    \caption{PET-OMAT model hyperparameters.}    
    \label{tab:pet-omat}
    \begin{tabular}{c cccc}
        \hline
        Model Size & $d_{\text{pet}}$ & $N_{\text{gnn}}$ & $N_{\text{tl}}$ & $r_{\text{cut}}$ (\AA) \\
        \hline
        XS & 128 & 2 & 1 & 4.0 \\
        S  & 256 & 3 & 1 & 4.5 \\
        M  & 384 & 3 & 2 & 4.5 \\
        L  & 512 & 4 & 2 & 4.5 \\
        \hline
    \end{tabular}
\end{table}

\begin{figure}[h!]
    \centering
    \includegraphics[width=\linewidth]{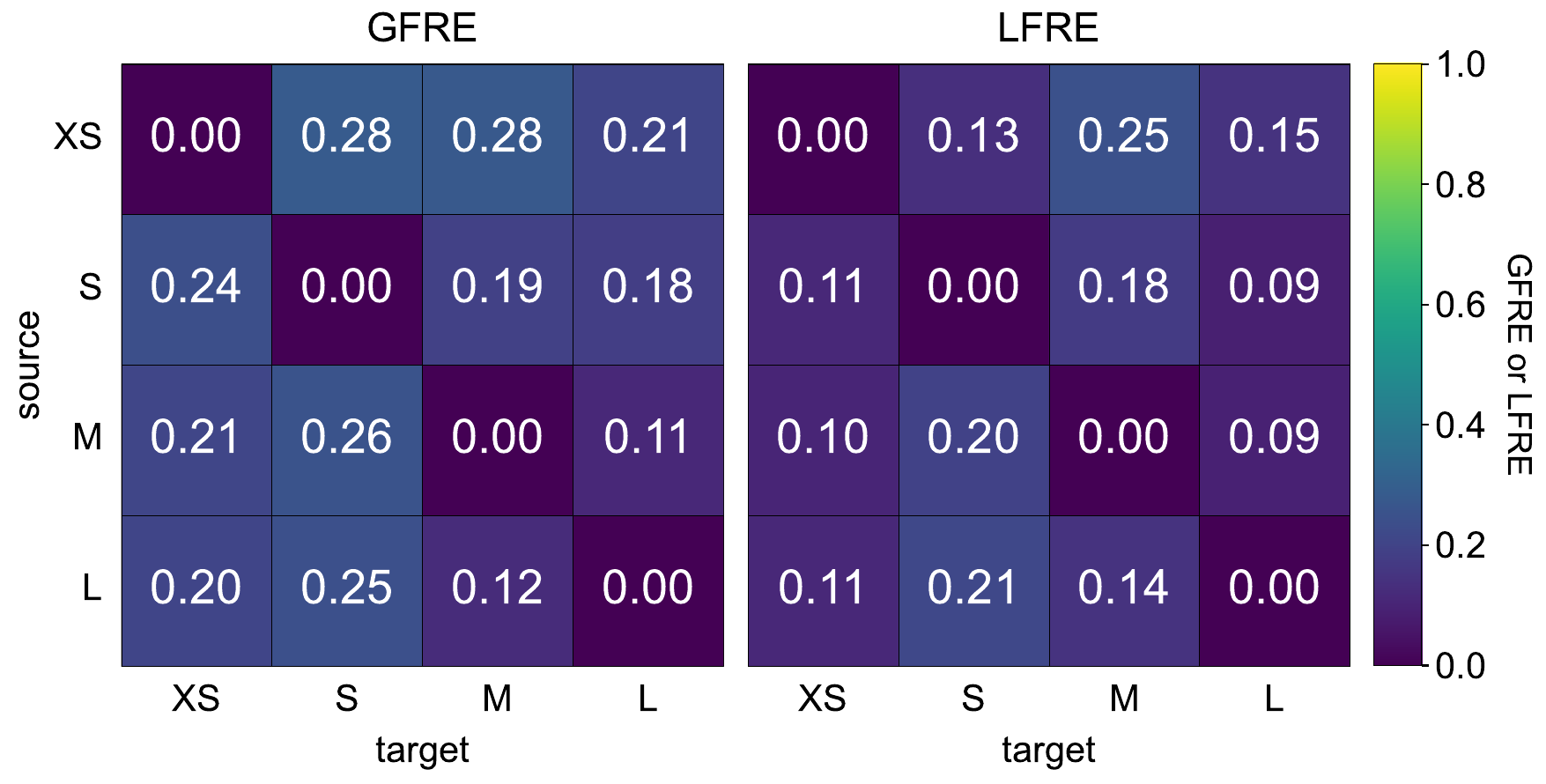}
    \caption{Reconstruction errors across the PET-OMAT checkpoints of different sizes computed for the MAD test subset.
    }
    \label{fig:pet-omat}
\end{figure}

From the error heat maps (see Fig.~\ref{fig:pet-omat}), the models seem to share a common backbone in the latent space: smaller models (XS, S) can capture the main latent feature manifold of larger models (M, L) with reasonable accuracy. Local reconstruction errors are higher and have stronger asymmetry, which indicates that local feature neighborhoods are progressively better reconstructed as the model size increases. We argue that the M model provides a favorable tradeoff, achieving low errors when reconstructing the L model, and can thus serve as an effective pre-training base for fine-tuning while remaining computationally efficient.

\subsection{Single vs.~multi-head architecture of the same model}
\label{sec:si-variants-mace-archictures}

We next compare the latent spaces of the single-head MACE-OMAT-0, the MACE architecture \cite{MACE} fully trained on OMat24 \cite{omat24} (MACE-OMAT-0), against the corresponding OMat24-specific readout head of the multi-head MACE-MH-1 model \cite{mace_multihead}. The multi-head variant is the MACE architecture enhanced with nonlinear interaction blocks and trained with multi-head replay fine-tuning, where task-specific readout heads are attached to a shared backbone and simultaneously trained on new data plus a replay buffer sampled from the original foundational training data to prevent catastrophic forgetting. As shown in Table \ref{tab:variants-mace-architectures}, reconstruction is nearly symmetric with moderate GFRE but very low LFRE. This hints that the multi-head design preserves the local representational fidelity while enhancing the global readout. In other words, the latent feature space remains essentially unchanged locally, but globally, the newly devised training strategy allows the multi-head model to alter the manifold.

\begin{table}[tbh]
    \centering
    \caption{GFRE and LFRE comparison between MACE-OMAT-0 (original MACE trained on OMAT) and MACE-MH-1-OMAT (OMAT head of the multi-head fine-tuned MACE model).}    
    \begin{tabular}{c cc}
        \hline
        Reconstruction direction (source  $\rightarrow$ target) & GFRE & LFRE \\
        \hline
        OMAT-0 $\rightarrow$ MH-1-OMAT & 0.32 & 0.01 \\
        MH-1-OMAT $\rightarrow$ OMAT-0 & 0.35 & 0.1 \\
        \hline
    \end{tabular}
    \label{tab:variants-mace-architectures}
\end{table}

\section{Varying targets}
\label{sec:si-dos}

\begin{table}[h!]
    \centering
    \caption{GFRE and LFRE comparison between PET-MAD and PET-MAD-DOS.}    
    \begin{tabular}{c cc}
        \hline
        Reconstruction direction (source  $\rightarrow$ target) & GFRE & LFRE \\
        \hline
        PET-MAD $\rightarrow$ PET-MAD-DOS & 0.68 & 0.39 \\
        PET-MAD-DOS $\rightarrow$ PET-MAD & 0.56 & 0.28 \\
        \hline
    \end{tabular}
    \label{tab:variants-dos}
\end{table}

Here, we evaluate the information content overlap between the last-layer features of PET-MAD (trained on energies and forces) and PET-MAD-DOS (trained on the electronic density of states) via feature reconstruction errors on the MAD test subset. The results show lower errors when reconstructing PET-MAD features from PET-MAD-DOS features compared to the reverse direction (Table~\ref{tab:variants-dos}). It indicates that the DOS-based features capture aspects sufficient for approximating energy-based features, while encoding additional information that the energy-only latent feature space may be oblivious to.

\section{Fine-tuning}
\label{sec:si-fine-tuning}

\subsection{Training errors and hyperparameters}
The errors of each PET checkpoints corresponding to the different fine-tuning strategies have been evaluated on the test split of the LPS dataset composed of 412 structures. Results are shown in Table~\ref{tab:lips-errors}. We also report the hyperparameters used for the PET-based models in Table~\ref{tab:lips-hypers}.

\begin{table*}[tbh]
    \centering
    \caption{Test set performance on LPS dataset for different PET-based models: PET-MAD, a pre-trained PET uMLIP baseline; BESPOKE, model trained from scratch; FF, model with full fine-tuning; HF, PET with head-only fine-tuning; FTL, model with full transfer learning training; HTL, model with head transfer learning.}    
    \begin{tabular}{c cccc}
        \hline
        Model & Energy RMSE & Energy MAE  & Forces RMSE & Forces MAE  \\
        & (meV/atom)& (meV/atom) &  (meV/Å) & (meV/Å)\\
        \hline
        PET-MAD & 210.01 & 208.26 & 129.69 & 75.95 \\
        BESPOKE & 4.19 & 2.44 & 99.71 & 51.23 \\
        FF & 8.73 & 6.58 & 89.44 & 54.59 \\
        HF & 8.57 & 6.98 & 96.30 & 58.28 \\
        FTL & 6.18 & 4.76 & 91.26 & 49.65 \\
        HTL & 8.70 & 7.17 & 98.10 & 58.44 \\
        \hline
    \end{tabular}
    \label{tab:lips-errors}
\end{table*}

\begin{table*}[tbh]
    \centering
    \caption{Hyperparameters used for PET-based models across training strategies on the LPS dataset.}    
    \begin{tabular}{c c}
        \hline
        Parameter & Value \\
        \hline
        Cutoff ($r_{\text{cut}}$) & 4.5 \\
        NN width ($d_{\text{pet}}$) & 128 \\
        Number of GNN layers ($N_{\text{gnn}}$) & 2 \\
        Number of transformer layers ($N_{\text{tl}}$) & 2 \\
        Number of heads ($N_{\text{heads}}$) & 8 \\
        Learning rate & $1\times10^{-5}$ \\
        Batch size & 16 \\
        \hline
    \end{tabular}
    \label{tab:lips-hypers}
\end{table*}

\subsection{Latent space projection}

In addition to the analyses in the main text, here we perform the PCovR analysis using the last-layer features extracted from PET checkpoints obtained with different fine-tuning techniques: full fine-tuning (FT), head fine-tuning (HF), full transfer learning (FTL), and head transfer learning (HTL) upon the LPS dataset. We also extract corresponding features from PET-MAD, which has been used as a pre-training basis. For the regression term of PCovR, we use the ground-truth cohesive energy, and we form the structure-level features by using the second order cumulant or the concatenation of the mean and standard deviation of the atomic feature vectors. We set the mixing parameter $\alpha=0.5$ to balance the regression and PCA terms, thereby enforcing consistency between checkpoint projections. Note that the PCovR space was first fitted using the embeddings from the MAD test subset, and the LPS projections were simply projected into this space.

The projections are consistent and similar across checkpoints, meaning that PET-derived representations retain the same global structure due to the common pre-trained backbone. We note that relatively larger fluctuations are observed for the transfer learning variants (FTL, HTL), which can be explained by the involvement of new heads with randomized initial weights. The similarities and differences observed in the PCovR projects are well-aligned with the global and local feature reconstruction errors presented in Section \ref{subsec:errors-umlip}.

\begin{figure}[tbh]
    \centering
    \includegraphics[width=\linewidth]{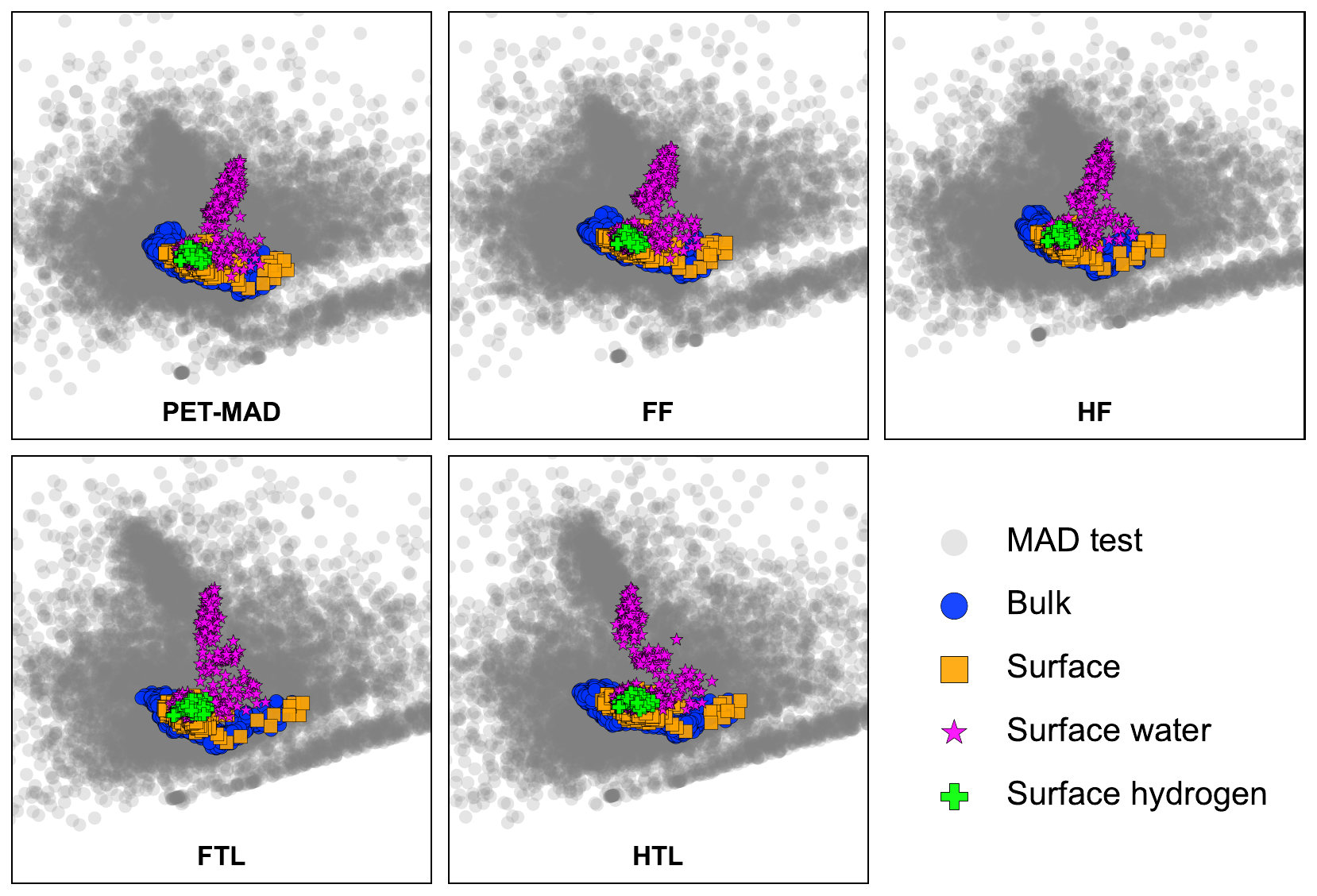}
    \caption{Two-dimensional PCovR projections of the LPS dataset constructed using the last-layer features of PET checkpoints trained with different fine-tuning setups. The greyscale background points correspond to the test subset of the MAD dataset, and the colored points in each panel correspond to the PCovR projections of the LPS systems ``within'' the MAD latent space.}
    \label{fig:lips-pcovr}
\end{figure}

\section{Backbone vs. last-layer features}
\label{sec:si-ll-vs-bb}

\begin{figure}[tpb]
    \centering
    \includegraphics[width=\linewidth]{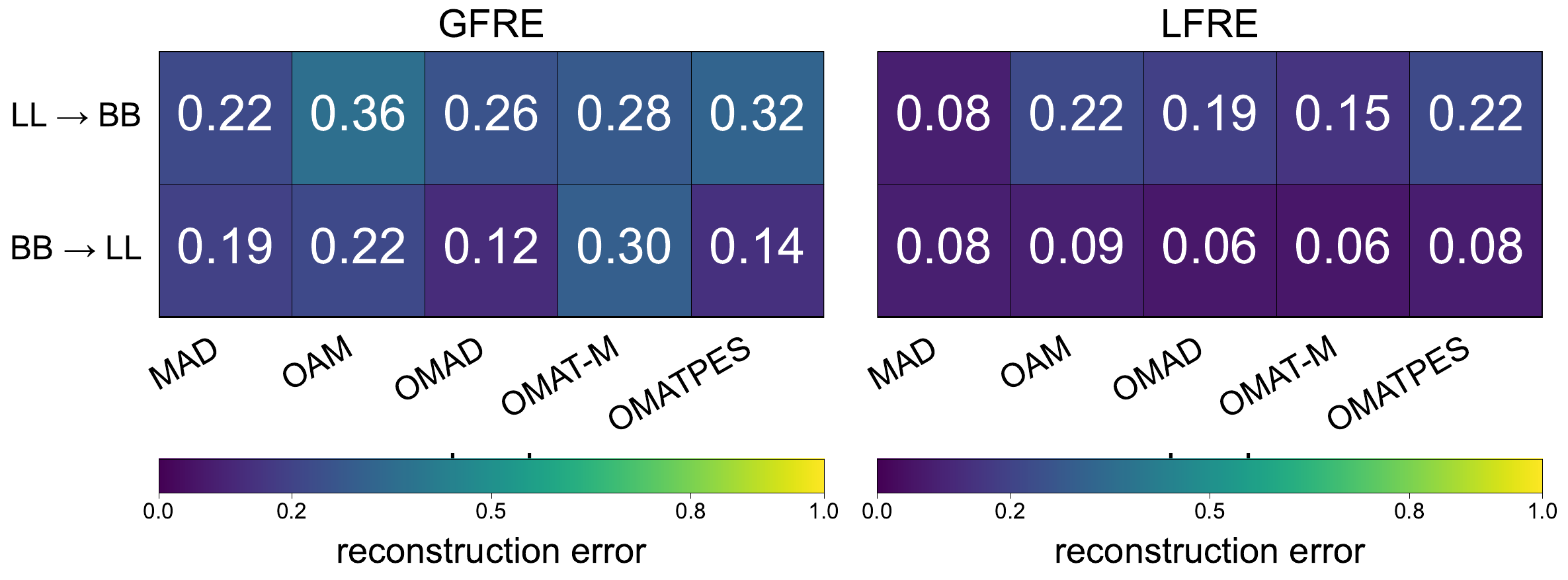}
    \caption{Global and local reconstruction errors for the last-layer features (LL) and backbone features (BB) of different PET checkpoints computed for the MAD test dataset.}
    \label{fig:ll_vs_bb-pet}
\end{figure}

We compare backbone (BB) and last-layer (LL) features across five PET variants, described in SI \ref{sec:si-variants-pet}. Checkpoints pre-trained on OMat24, corresponding to OAM, OMAD, OMAT, and OMATPES columns, show higher LL to BB reconstruction errors  (Fig. \ref{fig:ll_vs_bb-pet}) than the non-pre-trained MAD, which indicates that pre-training can enrich the BB features with transferable information. Overall, the flow is asymmetric, with an average GFREs of 0.29 for LL $\rightarrow$ BB and 0.19 for BB $\rightarrow$ LL, suggesting that BB features encode more information compared to the LL features. 

\section{Local to global features}
\label{sec:si-cumulants}

The cumulants $\boldsymbol{\kappa}^{(k)}_S$ are computed from the central moment $\boldsymbol{\mu}_{S}^{(k)}$ as follows:

\begin{align}
\boldsymbol{\kappa}^{(1)}_S &= \bar{\boldsymbol{\xi}}_S^{(F)}, \\
\boldsymbol{\kappa}^{(2)}_S &= \boldsymbol{\mu}^{(2)}_S, \\
\boldsymbol{\kappa}^{(3)}_S &= \boldsymbol{\mu}^{(3)}_S, \\
\boldsymbol{\kappa}^{(4)}_S &= \boldsymbol{\mu}^{(4)}_S - 3 (\boldsymbol{\mu}^{(2)}_S)^2, \\
\boldsymbol{\kappa}^{(5)}_S &= \boldsymbol{\mu}^{(5)}_S - 10 \boldsymbol{\mu}^{(2)}_S \boldsymbol{\mu}^{(3)}_S, \\
\boldsymbol{\kappa}^{(6)}_S &= \boldsymbol{\mu}^{(6)}_S - 15 \boldsymbol{\mu}^{(2)}_S \boldsymbol{\mu}^{(4)}_S - 10 (\boldsymbol{\mu}^{(3)}_S)^2 + 30 (\boldsymbol{\mu}^{(2)}_S)^3, \\
\boldsymbol{\kappa}^{(7)}_S &= \boldsymbol{\mu}^{(7)}_S - 21 \boldsymbol{\mu}^{(2)}_S \boldsymbol{\mu}^{(5)}_S - 35 \boldsymbol{\mu}^{(3)}_S \boldsymbol{\mu}^{(4)}_S + 210 \boldsymbol{\mu}^{(3)}_S (\boldsymbol{\mu}^{(2)}_S)^2, \\
\boldsymbol{\kappa}^{(8)}_S &= \boldsymbol{\mu}^{(8)}_S - 28 \boldsymbol{\mu}^{(2)}_S \boldsymbol{\mu}^{(6)}_S - 56 \boldsymbol{\mu}^{(3)}_S \boldsymbol{\mu}^{(5)}_S - 35 (\boldsymbol{\mu}^{(4)}_S)^2 \\
&\quad + 420 \boldsymbol{\mu}^{(4)}_S (\boldsymbol{\mu}^{(2)}_S)^2 + 560 (\boldsymbol{\mu}^{(3)}_S)^2 \boldsymbol{\mu}^{(2)}_S - 630 (\boldsymbol{\mu}^{(2)}_S)^4.
\end{align}

\begin{figure}[h!]
    \centering
    \includegraphics[width=\linewidth]{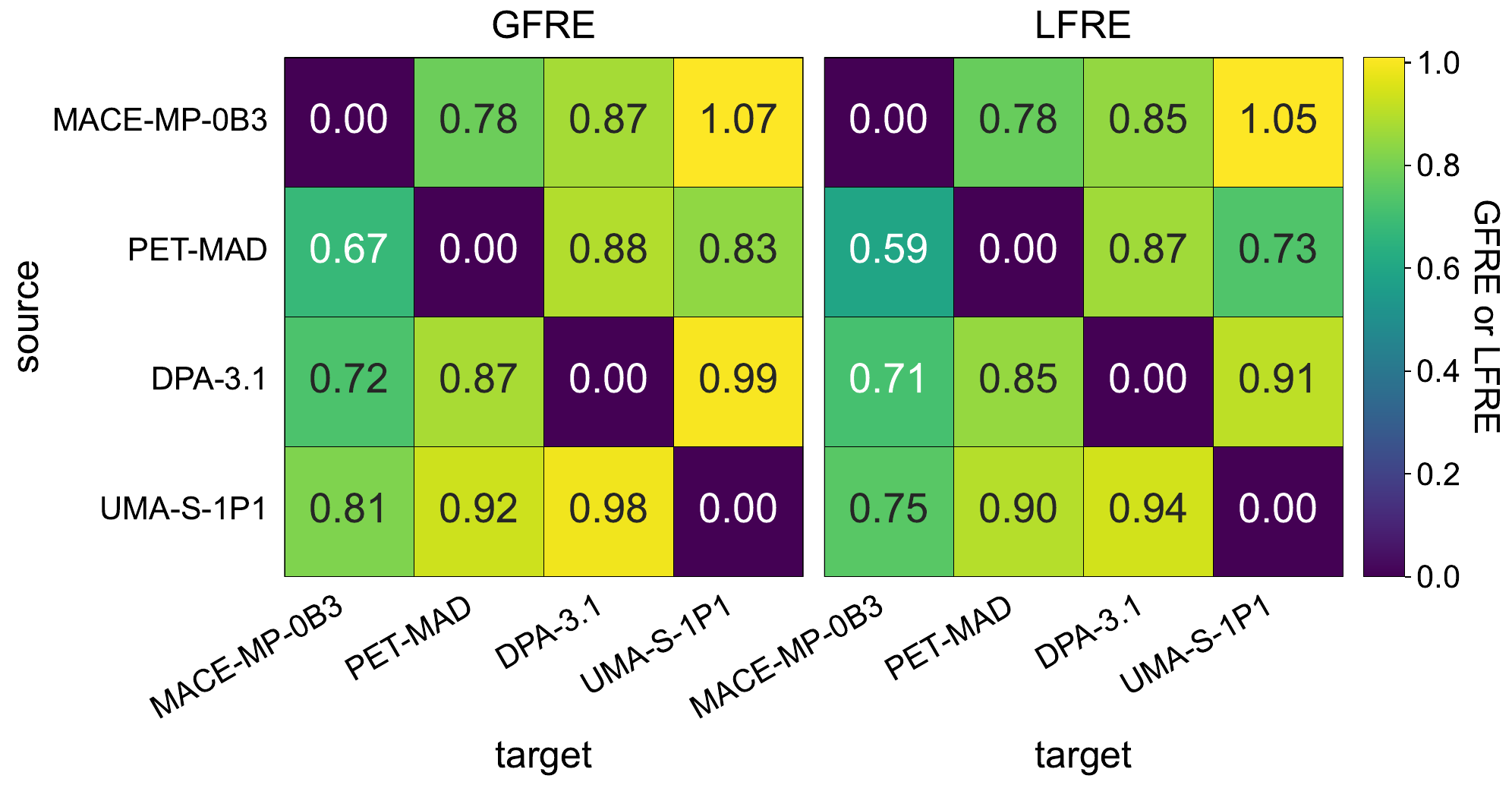}
    \caption{Reconstruction errors of the eight-order cumulative structure-level features across the uMLIPs on the MAD test dataset.}
    \label{fig:cumulants-errors-umlip}
\end{figure}

The global and local feature reconstruction errors across the uMLIPs computed with $\boldsymbol{\kappa}^{(8)}_S$ are presented in Fig.~\ref{fig:cumulants-errors-umlip}.

\end{document}